\documentstyle[12pt]{article}
\begin{document}

\title{Studies on Santilli's Locally Anisotropic and Inhomogeneous
 Isogeometries:\\
 I.\ Isobundles and Generalized Isofinsler Gravity }
\author{Sergiu I.\ Vacaru}
\date{{\small {\sl Institute of Applied Physics}\\
Academy of Sciences of Moldova\\
5 Academy str., Chi\c sin\u au 2028\\
{\sf Republic of Moldova}\\
\vskip6pt Fax: 011-3732-738149\\
E-mail: lises@cc.acad.md}}
\maketitle

\begin{abstract}
 We generalize the geometry of Santilli's locally anisotropic and inhomogeneous
 isospaces to the geometry of vector isobundles  provided with nonlinear and
 distinguished isoconnections and isometric  structures. We present,
 apparently for the first time, the isotopies of Lagrange,
Finsler and Kaluza--Klein
 spaces. We also continue the study of the
 interior, locally anisotropic and inhomogeneous gravitation by extending
 the isoriemannian space's constructions and presenting a geometric
 background  for the theory of isofield interactions  in
  generalized isolagrange and isofinsler spaces.
\end{abstract}
\vskip30pt
In press at Algebras, Groups and Geometries, Vol. 14, 1997

1991 Math. Subj. Class. 51N30, 51p05, 53-XX
\newpage
\section{Introduction}

A number of physical problems connected with the general interior dynamics
of deformable particles while moving within inhomogeneous and anisotropic
physical media result in a study of the most general known systems which are
nonlinear in coordinates $x$ and their derivatives $\dot{x}, \ddot{x},...,$
on wave functions and $\psi$ and their derivatives $\partial \psi ,
\partial\partial \psi ,... .$ Such systems are also nonlocal because of
possible integral dependencies on all of the proceeding quantities and
noncanonical with violation of integrability conditions for the existence of
a Lagrangian or a Hamiltonian \cite{Santilli19781985}.

The mathematical methods for a quantitative treatment of the latter
nonlinear, nonlocal and nonhamiltonian systems have been identified by Santilli
in a series of contributions beginning  the late 1970's \cite
{Santilli1978, Santilli1979, Santilli1996} under the name of isotopies, and
 include axiom preserving liftings of fields of numbers, vector and metric
 spaces, differential and integral calculus, algebras and geometries. These
 studies were then continued by a number of authors (see ref. \cite{Baltzer}
 for a comprehensive literature up to 1985, and monographs
 \cite{Kadeisvili,Lohmus,Santilli19781985,Santilli1979,Santilli1980,
TsagasSourlas} for subsequent literature).

This paper is devoted to a study of Santilli's isospaces and isogeometries
 over isofields treated via the isodifferential calculus according to their
latest formulation \cite{Santilli1980} (we extend this calculus
 for isospaces provided with nonlinear isoconnection structure). We shall
also use Kadeisvili's notion of isocontinuity \cite{Kadeisvili} and the
novel Santilli--Tsagas--Sourlas isodifferential topology \cite{Santilli1996,
TsagasSourlas}.

After reviewing the basic elements for completeness as well as for notational
convenience, we shall extend  Santilli's foundations of the isosympletic
 geometry \cite{Santilli1996} to isobundles and related aspects
 (by applying, in an isotopic manner, the methods summarized in
 Miron and Anastasiei  \cite{MironAnastasiei} and
 Yano and Ishihara \cite{YanoIshihara} monographs). We shall study,
 apparently for the first time, the isotopies of Lagrange, Finsler and
 Kaluza--Klein geometries. We shall then apply the results to further
 studies of the isogravitational theories (for isoriemannian spaces
 firstly considered by Santilli \cite{Santilli1996}) on vector isobundle
 provided with
 compatible nonlinear and distinguished isoconnections and isometric
 structures. Such  isogeometrical models of isofield interaction isotheories
 are in general nonlinear, nonlocal and
 nonhamiltonian and contain a very large class of local anisotropies
 and inhomogeneities induced  by four fundamental  isostructures:
 the partition of unity,  nonlinear isoconnection, distinguished
 isoconnections and isometric.

The novel geometric profile emerging from all the above studies is rather
 remarkable inasmuch as the first class of all isotopies herein
considered (called Kadeisvili's Class I \cite{Kadeisvili}) preserves
the abstract axioms of conventional formulations, yet permits a clear
broadening of their applicability, and actually result to be
''directly universal'' \cite{Santilli1996} for a number of possible well
 behaved nonlinear, nonlocal and nonhamiltonian systems. In turn, this
 permits a number of geometric unification such as that of all possible
 metrics (on isospaces with trivial nonlinear isoconnection structure)
 of a given dimension into Santilli's isoeuclidean metric, the
 unification of  exterior and interior gravitational problems despite
 their sizable structural differences and other unification.

The view adopted in this work is that a general field theory should
incorporate all possible anisotropic, inhomogeneous
 and stochastic manifestations of
classical and quantum interactions and, in consequence, corresponding
modifications of basic principles and mathematical methods have to be
introduced in order to formulate physical theories. There are established
three approaches for modeling  field interactions and spaces
anisotropies. The first one is to deal with a usual locally isotropic
physical theory and to consider anisotropies as a consequence of the
anisotropic structure of sources in field equations (for instance, a number
of cosmological models are proposed in the framework of the Einstein theory
with the energy--momentum generated by anisotropic matter, as a general
reference see \cite{MisnerThorneWheeler}). The second approach to
anisotropies originates from the Finsler geometry \cite{Finsler, Cartan,
Rund, Matsumoto} and its generalizations \cite{Asanov, MironAnastasiei,
AntonelliMiron, Bejancu, VacaruJMP, VacaruGoncharenko} with a general
imbedding into Kaluza--Klein (super) gravity and string theories \cite
{VacaruAP, VacaruNP, VacaruSupergravity}, and speculates a generic
anisotropy of the space--time structure and of fundamental field of
interactions. The Santilli's approach is more radical by proposing a
generalization of Lie theory and introducing isofields, isodualities and
related mathematical structures. Roughly speaking, by using corresponding
partitions of the unit we can model possible metric anisotropies as in
Finsler or generalized Lagrange geometry but the problem is also to take
into account classes of anisotropies generated by nonlinear and
distinguished connections.

This is a first paper from a series of works devoted to the formulation of
the theory of inhomogeneous,
locally anisotropic and higher order anisotropic isofield
interactions (here we note that the term ''higher order'' is used as a
 general one for higher order tangent bundles \cite{YanoIshihara}, or
 higher order extensions of vector superbundles \cite{VacaruNP,
VacaruSupergravity}, in a number of lines
 alternative to jet bundles \cite{Saunders,Sardanashvily}, and only
 under corresponding constraints one obtains the geometry of higher order
 Lagrangians \cite{MironAtanasiu}).
 The main purpose of this paper is to present a synthesis of
the Santilli isotheory and the approach on modeling locally anisotropic
geometries and physical models on bundle spaces provided with nonlinear
connection and distinguished connection and metric structures
 \cite{MironAnastasiei,YanoIshihara}.
 The isotopic
variants of generalized Lagrange and Finsler geometry will be analyzed.
Basic geometric constructions such as nonlinear isoconnections in vector
isobundles, the isotopic curvatures and torsions of distinguished isoconnections
and theirs structure equations and invariant values will be defined. A model
of locally anisotropic and inhomogeneous
gravitational isotheory will be constructed.

Section 2 is devoted to basic notations and definitions on Santilli and
coauthors isotheory. We introduce the bundle isospaces in Section 3 where
some necessary properties of Lie--Santilli isoalgebras and isogroups and
corresponding isotopic extensions of manifolds are applied in order to
define fiber isospaces and consider their such (being very important for
modeling of isofield interactions) classes of principal isobundles and
vector isobundles. The isogeometry of nonlinear isoconnections in vector
 isobundles is studied in Section 4. Isotopic distinguishing of geometric
objects, the isocurvatures and isotorsions of nonlinear and distinguished
 isoconnections, the
structure equations and invariant conditions are defined in Section 5. The
next Section 6 is devoted to the isotopic extensions of generalized Lagrange
and Finsler geometries. The isofield equations of locally anisotropic and
inhomogeneous interactions will be analyzed in Section 7. The outlook and
conclusions are contained in Section 8.

\section{Basic Notions on Isotopies}

In this section we shall mainly recall some necessary fundamental notions
and refer to works \cite{Santilli1996, Kadeisvili} for details and
references on Lie--Santilli isotheory.

\subsection{Isotopies of the unit and isospaces}

The {\bf isotheory} is based on the concept of fundamental {\bf isotopy}
which is the lifting $I\rightarrow \widehat{I}$ of the $n$--dimensional unit
$I=diag\left( 1,1,...,1\right) $ of the Lie's theory into an $n\times n$%
--dimensional matrix
$$
\widehat{I}=\left( I_j^i\right) =\widehat{I}\left( t,x,\dot x,\ddot x,\psi
,\psi ^{+},\partial \psi ,\partial \psi ^{+},\partial \partial \psi
,\partial \partial \psi ^{+},...\right)
$$
called the isounit. For simplicity, we consider that maps $I\rightarrow
\widehat{I}$ are of necessary Kadeisvili Class I (II), the Class III being
considered as the union of the first two, i. e. they are sufficiently
smooth, bounded, nowhere degenerate, Hermitian and positive (negative)
definite, characterizing isotopies (isodualities).

One demands a compatible lifting of all associative products $AB$ of some
generic quantities $A$ and $B$ into the isoproduct $A*B$ satisfying the
properties:%
$$
AB\Rightarrow A*B=A\widehat{T}B,~I A=A I \equiv
A\rightarrow \widehat{I}*A=A*\widehat{I}\equiv A,
$$
$$
A\left( BC\right) =\left( AB\right) C\rightarrow A*\left( B*C\right) =\left(
A*B\right) *C,
$$
where the fixed and invertible matrix $\widehat{T}$ is called the isotopic
element.

To follow our outline, a conventional field $F\left( a,+,\times \right) ,$
for instance of real, complex or quaternion numbers, with elements $a$,
conventional sum + and product $a\times b\doteq ab,\,$ must be lifted into
the so--called isofield $\widehat{F}\left( \widehat{a},+,*\right) ,$
satisfying properties%
$$
F\left( a,+,*\right) \rightarrow \widehat{F}\left( \widehat{a},+,*\right) ,~%
\widehat{a}=a\widehat{I}
$$
$$
\widehat{a}*\widehat{b}=\widehat{a}\widehat{T}\widehat{b}=\left( ab\right)
\widehat{I},~\widehat{I}=\widehat{T}^{-1}
$$
with elements $\widehat{a}$ called isonumbers, $+$ and $*$ are conventional
sum and isoproduct preserving the axioms of the former field $F\left(
a,+,\times \right) .$ All operations in $F$ are generalized for $\widehat{F}%
, $ for instance we have isosquares $\widehat{a}^{\widehat{2}}=\widehat{a}*%
\widehat{a}=\widehat{A}\widehat{T}\widehat{a}=a^2\widehat{I},$ isoquotient $%
\widehat{a}\widehat{/}\widehat{b}=\left( a/b\right) \widehat{I},$ isosquare
roots $\widehat{a}^{1/2}=a^{1/2}\widehat{I},...;$\ $\widehat{a}*A\equiv
aA.\, $ We note that in the literature one uses two types of denotation for
isotopic product $*$ or $\widehat{\times }$ (in our work we shall consider $%
*\equiv \widehat{\times }).$

Let us consider, for example, the main lines of the isotopies of a $n$%
--di\-men\-si\-o\-nal Euclidean space $E^n\left( x,g,{\cal R}\right) ,$
where ${\cal R}\left( n,+,\times \right) $ is the real number field,
provided with a local coordinate chart $x=\{x^k\},k=1,2,...,n,$ and $n$%
--dimensional metric $\rho =\left( \rho _{ij}\right) =diag\left(
1,1,...,1\right) .$ The scalar product of two vectors $x,y\in E^n$ is
defined as
$$
\left( x-y\right) ^2=\left( x^i-y^i\right) \rho _{ij}\left( x^j-y^j\right)
\in {\cal R}\left( n,+,\times \right)
$$
were the Einstein summation rule on repeated indices is assumed hereon.

The {\bf Santilli's isoeuclidean} spaces
 $\widehat{E}\left( \widehat{x},\widehat{\rho },%
\widehat{R}\right) $ of Class III are introduced as $n$--dimensional metric
spaces defined over an isoreal isofield $\widehat{R}\left( \widehat{n},+,%
\widehat{\times }\right) $ with an $n\times n$--dimensional real--valued and
symmetrical isounit $\widehat{I}=\widehat{I}^t$ of the same class, equipped
with the ''isometric''%
$$
\widehat{\rho }\left( t,x,v,a,\mu ,\tau ,...\right) =\left( \widehat{\rho }%
_{ij}\right) =\widehat{T}\left( t,x,v,a,\mu ,\tau ,...\right) \times \rho =%
\widehat{\rho }^t,
$$
where $\widehat{I}=\widehat{T}^{-1}=\widehat{I}^t.$

A local coordinate cart on $\widehat{E}\left( \widehat{x},\widehat{\rho },%
\widehat{R}\right) $ can be defined in contravariant
$$
\widehat{x}=\{\widehat{x}^k=x^{\widehat{k}}\}=\{x^k\times \widehat{I}_{~k}^{%
\widehat{k}}\}
$$
or covariant form%
$$
\widehat{x}_k=\widehat{\rho }_{kl}\widehat{x}^l=\widehat{T}_k^r\rho
_{ri}x^i\times \widehat{I},
$$
where $x^k,x_k\in \widehat{E}$. The square of ''isoeuclidean distance''
between two points $\widehat{x},\widehat{y}\in \widehat{E}$ is defined as
$$
\left( \widehat{x}-\widehat{y}\right) ^{\widehat{2}}=\left[ \left( \widehat{x%
}^i-\widehat{y}^i\right) \times \widehat{\rho }_{ij}\times \left( \widehat{x}%
^j-\widehat{y}^j\right) \right] \times \widehat{I}\in \widehat{R}
$$
and the isomultiplication is given by
$$
\widehat{x}^{\widehat{2}}=\widehat{x}^k\widehat{\times }\widehat{x}_k=\left(
x^k\times \widehat{I}\right) \times \widehat{T}\times \left( x_k\times
\widehat{I}\right) =\left( x^k\times x_k\right) \times \widehat{I}=n\times
\widehat{I} .
$$

Whenever confusion does not arise isospaces can be practically treated via
the conventional coordinates $x^k$ rather than the isotopic ones $\widehat{x}%
^k=x^k\times \widehat{I}.$ The symbols $x,v,a,...$ will be used for
conventional spaces while symbols $\widehat{x},\widehat{v},\widehat{a},...$
will be used for isospaces; the letter $\widehat{\rho }\left(
x,v,a,...\right) $ refers to the projection of the isometric $\widehat{\rho
}$ in the original space.

We note that an isofield of Class III, explicitly denoted as $\widehat{F}%
_{III}\left( \widehat{a},+,\widehat{\times }\right) $ is a union of two
disjoint isofields, one of Class I, $\widehat{F}_I\left( \widehat{a},+,%
\widehat{\times }\right) ,$ in which the isounit is positive definite, and
one of Class II, $\widehat{F}_{II}\left( \widehat{a},+,\widehat{\times }%
\right) ,$ in which the isounit is negative--definite. The Class II\ of
isofields is usually written as $\widehat{F}^d\left( \widehat{a}^d,+,%
\widehat{\times }^d\right) $ and called isodual fields with isodual unit $%
\widehat{I}^d=-\widehat{I}<0,$ isodual isonumbers $\widehat{a}^d=a\times
\widehat{I}^d=-\widehat{a},$ isodual isoproduct $\widehat{\times }^d=\times
\widehat{T}^d\times =-\widehat{\times },$ etc.
For simplicity, in our further considerations we shall use the general
 terms isofields, isonumbers even for isodual fields, isodual numbers and
so on if this will not give rise to ambiguities.

\subsection{Isocontinuity and isotopology}

The isonorm of an isofield of Class III is defined as%
$$
\left\uparrow \widehat{a}\right\uparrow =\left| a\right| \times \widehat{I}
$$
where $\left| a\right| $ is the conventional norm. Having defined a function
$\widehat{f}\left( \widehat{x}\right) $ on isospace $\widehat{E}\left(
\widehat{x},\widehat{\delta },\widehat{R}\right) $ over isofield $\widehat{R}%
\left( \widehat{n},+,\widehat{\times }\right) $ one introduces (see details
and references in \cite{Kadeisvili}) the isomodulus
$$
\left\uparrow \widehat{f}\left( \widehat{x}\right) \right\uparrow =\left|
\widehat{f}\left( \widehat{x}\right) \right| \times \widehat{I}
$$
where $\left| \widehat{f}\left( \widehat{x}\right) \right| $ is the
conventional modulus.

One says that an infinite sequence of isofunctions of Class I\ $\widehat{f}%
_1,\widehat{f}_2,...$ is ''strongly isoconvergent'' to the isofunction $%
\widehat{f}$ of the same class if
$$
\lim _{k\rightarrow \infty }\left\uparrow \widehat{f}_k-\widehat{f}%
\right\uparrow =\widehat{0}.
$$
The Cauchy isocondition is expressed as
$$
\left\uparrow \widehat{f}_m-\widehat{f}_n\right\uparrow <\widehat{\rho }%
=\rho \times \widehat{I}
$$
where $\delta $ is real and $m$ and $n$ are greater than a suitably chosen $%
N\left( \rho \right) .$ Now the isotopic variants of continuity, limits,
series, etc, can be easily constructed in a traditional manner.

The notion of $n$--dimensional isomanifold was studied by Tsagas and Sourlas
(we refer the reader for details in \cite{TsagasSourlas}). Their
constructions are based on idea that every isounit of Class III can always
be diagonalized into the form
$$
\widehat{I}=diag\left( B_1,B_2,...,B_n\right) ,B_k\left( x,...\right) \neq
0,k=1,2,...,n .
$$
In result of this one defines an isotopology $\widehat{\tau }$ on $\widehat{R%
}^n$ which coincides everywhere with the conventional topology $\tau $ on $%
R^n$ except at the isounit $\widehat{I}.$ In particular, $\widehat{\tau }$
is everywhere local--differential, except at $\widehat{I}$ which can
incorporate integral terms. The above structure is called the
Tsagas--Sourlas isotopology or an integro--differential topology.
Finally, in this subsection, we note that Prof. Tsagas and Sourlas used a
 conventional topology on isomanifolds. The isotopology was first introduced
 by Prof. Santilli in ref. \cite{Santilli1996}.

\subsection{Isodifferential and isointegral calculus}

Now we are able to introduce isotopies of the ordinary differential
calculus, i.e. the isodifferential calculus (for short).

The {\bf isodifferentials} of Class I of the contravariant and covariant
coordinates $\widehat{x}^k=x^{\widehat{k}}$ and $\widehat{x}_k=x_{\widehat{k}%
}$ on an isoeuclidean space $\widehat{E}$ of the same class is given by
$$
\widehat{d}\widehat{x}^k=\widehat{I}_{~i}^k\left( x,...\right) dx^i,~%
\widehat{d}\widehat{x}_k=\widehat{T}_k^{~i}\left( x,...\right) dx_i\eqno(2.1)
$$
where $\widehat{d}\widehat{x}^k$ and $\widehat{d}\widehat{x}_k$ are defined
on $\widehat{E}$ while the $\widehat{I}_{~i}^kdx^i$ and $\widehat{T}%
_k^{~~i}dx_i$ are the projections on the conventional Euclidean space.

For a sufficiently smooth isofunction $\widehat{f}\left( \widehat{x}\right) $
on a closed domain $\widehat{U}\left( \widehat{x}^k\right) $ covered by
contravariant iso\-co\-or\-di\-na\-tes $\widehat{x}^k$ we can define the
partial iso\-de\-ri\-va\-ti\-ves $\widehat{\partial }_{\widehat{k}}=\frac{%
\widehat{\partial }}{\widehat{\partial }\widehat{x}^k}$ at a point $\widehat{%
x}_{(0)}^k\in \widehat{U}\left( \widehat{x}^k\right) $ by considering the
limit%
$$
\widehat{f^{\prime }}\left( \widehat{x}_{(0)}^k\right) =\widehat{\partial }_{%
\widehat{k}}\widehat{f}\left( \widehat{x}\right) \mid _{\widehat{x}_{(0)}^k}=%
\frac{\widehat{\partial }\widehat{f}\left( \widehat{x}\right) }{\widehat{%
\partial }\widehat{x}^k}\mid _{\widehat{x}_{(0)}^k}=\widehat{T}_k^{~i}\frac{%
\partial f\left( x\right) }{\partial x^i}\mid _{\widehat{x}_{(0)}^k}=%
\eqno(2.2)
$$
$$
\lim _{\widehat{d}\widehat{x}^k\rightarrow \widehat{0}^k}\frac{\widehat{f}%
\left( \widehat{x}_{(0)}^k+\widehat{d}\widehat{x}^k\right) -\widehat{f}%
\left( \widehat{x}_{(0)}^k\right) }{\widehat{d}x^k}
$$
where $\widehat{\partial }\widehat{f}\left( \widehat{x}\right) /\widehat{%
\partial }\widehat{x}^k$ is computed on $\widehat{E}$ and $\widehat{T}%
_k^{~i}\partial f\left( x\right) /\partial x^i$ is the projection in $E.$

In a similar manner we can define the{\bf \ partial isoderivatives} $%
\widehat{\partial }^{\widehat{k}}=\frac{\widehat{\partial }}{\widehat{%
\partial }\widehat{x}_k}$ with respect to a covariant variable $\widehat{x}%
_k:$%
$$
\widehat{f^{\prime }}\left( \widehat{x}_{k(0)}\right) =\widehat{\partial }^{%
\widehat{k}}\widehat{f}\left( \widehat{x}\right) \mid _{\widehat{x}_{k(0)}}=%
\frac{\widehat{\partial }\widehat{f}\left( \widehat{x}\right) }{\widehat{%
\partial }\widehat{x}_k}\mid _{\widehat{x}_{k(0)}}=\widehat{T}_k^{~i}\frac{%
\partial f\left( x\right) }{\partial x^i}\mid _{\widehat{x}_{k(0)}}=%
\eqno(2.3)
$$
$$
\lim _{\widehat{d}\widehat{x}_{\widehat{k}}\rightarrow \widehat{0}_k}\frac{%
\widehat{f}\left( \widehat{x}_{k(0)}+\widehat{d}\widehat{x}_k\right) -%
\widehat{f}\left( \widehat{x}_{k(0)}\right) }{\widehat{d}x_k}.
$$

The isodifferentials of an isofunction of contravariant or covariant
coordinates, $\widehat{x}^k$ or $\widehat{x}_k,$ are defined according the
formulas%
$$
\widehat{d}\widehat{f}\left( \widehat{x}\right) \mid _{contrav}=\widehat{%
\partial }_{\widehat{k}}\widehat{f}\widehat{d}\widehat{x}^k=\widehat{T}%
_k^{~i}\frac{\partial f\left( x\right) }{\partial x^i}\widehat{I}_{~j}^kdx^j=%
\frac{\partial f\left( x\right) }{\partial x^k}dx^k=\frac{\partial f\left(
x\right) }{\partial x^i}\widehat{T}_j^{~i}dx^j
$$
and
$$
\widehat{d}\widehat{f}\left( \widehat{x}\right) \mid _{covar}=\widehat{%
\partial }^{\widehat{k}}\widehat{f}\widehat{d}\widehat{x}_k=\widehat{I}%
_{~i}^k\frac{\partial f\left( x\right) }{\partial x_i}\widehat{T}_k^{~j}dx_j=%
\frac{\partial f\left( x\right) }{\partial x_k}dx_k=\frac{\partial f}{%
\partial x_j}\widehat{I}_{~j}^idx_i.
$$

The second order isoderivatives there are introduced by iteration of the
notion of isoderivative:%
$$
\widehat{\partial }_{\widehat{i}\widehat{j}}^2\widehat{f}(\widehat{x})=\frac{%
\widehat{\partial }^2\widehat{f}(\widehat{x})}{\widehat{\partial }\widehat{x}%
^i\widehat{\partial }\widehat{x}^j}=\widehat{T}_{\widehat{i}}^{~i}\widehat{T}%
_{\widehat{j}}^{~j}\frac{\partial ^2f\left( x\right) }{\partial x^i\partial
x^j},
$$
$$
\widehat{\partial }^{2\widehat{i}\widehat{j}}\widehat{f}(\widehat{x})=\frac{%
\widehat{\partial }^2\widehat{f}(\widehat{x})}{\widehat{\partial }\widehat{x}%
_i\widehat{\partial }\widehat{x}_j}=\widehat{I}_{~i}^{\widehat{i}}\widehat{I}%
_{~j}^{\widehat{j}}\frac{\partial ^2f\left( x\right) }{\partial x_i\partial
x_j},
$$
$$
\widehat{\partial }_{\widehat{i}}^{2~\widehat{j}}\widehat{f}(\widehat{x})=%
\frac{\widehat{\partial }^2\widehat{f}(\widehat{x})}{\widehat{\partial }%
\widehat{x}^i\widehat{\partial }\widehat{x}_{\widehat{j}}}=\widehat{T}_{%
\widehat{i}}^{~i}\widehat{I}_j^{~\widehat{j}}\frac{\partial ^2f\left(
x\right) }{\partial x^i\partial x_j}.
$$

The Laplace isooperator on Euclidean space $\widehat{E}\left( \widehat{x},%
\widehat{\delta },\widehat{R}\right) $ is given by
$$
\widehat{\Delta }=\widehat{\partial }_k\widehat{\partial }^k=\widehat{%
\partial }^i\rho _{ij}\widehat{\partial }^j=\widehat{I}_{~k}^{\widehat{i}%
}\partial ^k\rho _{ij}\partial ^j\eqno(2.4)
$$
where there are also used usual partial derivatives $\partial ^j=\partial
/\partial x_j$ and $\partial _k=\partial /\partial x^k.$

The isodual isodifferential calculus is characterized by the following
isodual differentials and isodual isoderivatives%
$$
\widehat{d}^{(d)}\widehat{x}^{(d)k}=\widehat{I}_{\quad i}^{(d)k}d\widehat{x}%
^{(d)i}\equiv \widehat{d}x^k,\quad \widehat{\partial }^{(d)}/\widehat{%
\partial }\widehat{x}^{(d)i}\widehat{T}_k^{~i(d)}\partial /\partial \widehat{%
x}^{i(d)}\equiv \widehat{T}_k^{~i}\partial /\partial \widehat{x}^i.
$$

The formula (2.4) is different from the expression for the Laplace operator
$$
\Delta =\widehat{\rho }^{-1/2}\partial _i\widehat{\rho }^{1/2}\widehat{\rho }%
^{ij}\partial _j
$$
even though the Euclidean isometric $\widehat{\rho }\left( x,v,a,...\right) $
is more general than the Riemannian metric $g\left( x\right) .$ For partial
isoderivations one follows the next properties:%
$$
\frac{\widehat{\partial }\widehat{x}^i}{\widehat{\partial }\widehat{x}^j}%
=\delta _{~j}^i,~\frac{\widehat{\partial }\widehat{x}_i}{\widehat{\partial }%
\widehat{x}_j}=\delta _i^{~j},~\frac{\widehat{\partial }\widehat{x}_i}{%
\widehat{\partial }\widehat{x}^j}=\widehat{T}_i^{~j},~\frac{\widehat{%
\partial }\widehat{x}^i}{\widehat{\partial }\widehat{x}_j}=\widehat{I}%
_{~j}^i.
$$

Here we remark that isointegration (the inverse to isodifferential) is
defined \cite{Santilli1996} as to satisfy conditions
$$
\int^{\symbol{94}}\widehat{d}\widehat{x}=\int \widehat{T}\widehat{I}dx=\int
dx=x,
$$
where $\int^\symbol{94}=\int \widehat{T}.$

\subsection{Santilli's isoriemannian isospaces}

Let consider ${\cal R=R}\left( x,g,{\sl R}\right) $ a (pseudo)\ Riemannian
space over the reals ${\sl R}\left( n,+,\times \right) $ with local
coordinates $x=\{x^\mu \}$ and nonwere singular, symmetrical and real--valued
metric $g\left( x\right) =\left( g_{\mu \nu }\right) =g^t$ and the tangent
flat space $M\left( x,\eta ,{\sl R}\right) $ provided with flat real metric $%
\eta $ (for a corresponding signature and dimension we can consider $M$ as
the well known Minkowski space). The metric properties of the Riemannian
spaces are defined by scalar square of a tangent vector $x,$%
$$
x^2=x^\mu g_{\mu \nu }\left( x\right) x^v\in {\sl R}
$$
or, in infinitesimal form by the line element%
$$
ds^2=dx^\mu g_{\mu \nu }\left( x\right) dx^\nu
$$
and related formalism of covariant derivation (see for instance \cite
{MisnerThorneWheeler}).

The isotopies of the Riemannian spaces and geometry,
 were first studied and applied by \cite{Santilli1996} and
 are called Santilli's isoriemannian spaces and geometry.
 In this section we consider isoriemannian spaces equipped with
  the Santilli--Tsagas--Sourlas isotopology
 \cite{Santilli1996,TsagasSourlas} in a similar manner as we have done
in the previous subsection for isoeuclidean spaces but with respect to a
general, non flat, isometric. A isoriemannian space $\widehat{{\cal R}}{\cal %
=}\widehat{{\cal R}}\left( \widehat{x},\widehat{g},\widehat{{\sl R}}\right) ,
$ over the isoreals $\widehat{R}=\widehat{R}\left( \widehat{n},+,\widehat{x}%
\right) $ with common isounits $\widehat{I}=\left( \widehat{I}_{~\nu }^\mu
\right) =\widehat{T}^{-1},$ is provided with local isocoordinates $\widehat{x%
}=\{\widehat{x}^\mu \}=\{x^\mu \}$ and isometric $\widehat{g}\left(
x,v,a,\mu ,\tau ,...\right) =\widehat{T}\left( x,v,\mu ,\tau ,...\right)
g\left( x\right) ,$ where $\widehat{T}=\left( \widehat{T}_\mu ^{~\nu
}\right) $ is nowhere singular, real valued and symmetrical matrix of Class I
with $C^\infty $ elements. The corresponding isoline and infinitesimal
elements are written as
$$
\widehat{x}^{\widehat{2}}=[\widehat{x}^\mu \widehat{g}_{\mu \nu }\left(
x,v,a,\mu ,\tau ,...\right) \widehat{x}^\nu ]\times \widehat{I}\in \widehat{R%
}
$$
with infinitesimal version
$$
d\widehat{s}^2=(\widehat{d}\widehat{x}^\mu \widehat{g}_{\mu \nu }\left(
x\right) \widehat{d}\widehat{x}^\nu )\times \widehat{I}\in \widehat{R}.
$$

The {\bf covariant isodifferential calculus} has been introduced in ref.
 \cite{Santilli1996} via the expression
$$
\widehat{D}\widehat{X}^\beta =\widehat{d}\widehat{X}^\beta +\widehat{\Gamma }%
_{\alpha \gamma }^\beta \widehat{X}^\alpha \widehat{d}\widehat{x}^\gamma
$$
with corresponding covariant isoderivative%
$$
\widehat{X}_{\uparrow \mu }^\beta =\widehat{\partial }_\mu \widehat{X}^\beta
+\widehat{\Gamma }_{\alpha \mu }^\beta \widehat{X}^\alpha
$$
with the {\bf isocristoffel symbols} written as
$$
\widehat{\{\alpha \beta \gamma \}}=\frac 12\left( \widehat{\partial }_\alpha
\widehat{g}_{\beta \gamma }+\widehat{\partial }_\gamma \widehat{g}_{\alpha
\beta }-\widehat{\partial }_\beta \widehat{g}_{\alpha \gamma }\right) =%
\widehat{\{\gamma \beta \alpha \}},\eqno(2.5)
$$
$$
\widehat{\Gamma }_{\alpha \gamma }^\beta =\widehat{g}^{\beta \rho }\widehat{%
\{\alpha \rho \gamma \}}=\widehat{\Gamma }_{\alpha \gamma }^\beta ,
$$
where $\widehat{g}^{\beta \rho }$ is inverse to $\widehat{g}_{\alpha \beta }.
$

The crucial difference between Riemannian spaces and
iso\-spa\-ces is obvious if the corresponding auto--parallel equations
$$
\frac{Dx_\beta }{Ds}=\frac{dv_\beta }{ds}+\{\alpha \beta \gamma \}\left(
x\right) \frac{dx^\alpha }{ds}\frac{dx^\gamma }{ds}=0\eqno(2.6)
$$
and auto--isoparallel equations
$$
\frac{\widehat{D}\widehat{x}_\beta }{\widehat{D}\widehat{s}}=\frac{\widehat{d%
}v_\beta }{\widehat{d}\widehat{s}}+\widehat{\{\alpha \beta \gamma \}}\left(
\widehat{x},\widehat{v},\widehat{a},...\right) \frac{\widehat{d}\widehat{x}%
^\alpha }{\widehat{d}\widehat{s}}\frac{\widehat{d}\widehat{x}^\gamma }{%
\widehat{d}\widehat{s}}=0\eqno(2.7)
$$
where $\widehat{v}=\widehat{d}\widehat{x}/\widehat{d}\widehat{s}=\widehat{I}%
_S\times dx/ds,\widehat{s}$ is the proper isotime and $\widehat{I}_S$ is the
related one--dimensional isounit, can be identified by observing that
equations (2.6) are at most quadratic in the velocities while the isotopic
equations (2.7) are arbitrary nonlinear in the velocities and another
possible variables and parameters $\left( \widehat{a},...\right) .$

By using coefficients $\widehat{\Gamma }_{\alpha \gamma }^\beta $ we
introduce the next isotopic values \cite{Santilli1996}:

the {\bf isocurvature tensor}
$$
\widehat{R}_{\alpha ~\gamma \delta }^{\quad \beta }=\widehat{\partial }%
_\delta \widehat{\Gamma }_{\alpha \gamma }^\beta -\widehat{\partial }_\gamma
\widehat{\Gamma }_{\alpha \delta }^\beta +\widehat{\Gamma }_{\varepsilon
\delta }^\beta \widehat{\Gamma }_{\alpha \gamma }^\varepsilon -\widehat{%
\Gamma }_{\varepsilon \gamma }^\beta \widehat{\Gamma }_{\alpha \delta
}^\varepsilon ;\eqno(2.8)
$$

the {\bf isoricci tensor} $\widehat{R}_{\alpha \gamma }=\widehat{R}_{\alpha
~\gamma \beta }^{\quad \beta };$

the {\bf isocurvature scalar} $\widehat{R}=%
\widehat{g}^{\alpha \gamma }\widehat{R}_{\alpha \gamma };$

the {\bf isoeinstein tensor}
$$
\widehat{G}_{\mu \nu }=\widehat{R}_{\mu \nu }-\frac 12\widehat{g}_{\mu \nu }%
\widehat{R}\eqno(2.9)
$$

and the {\bf istopic isoscalar}%
$$
\widehat{\Theta }=\widehat{g}^{\alpha \beta }\widehat{g}^{\gamma \delta
}\left( \widehat{\{\rho \alpha \delta \}}\widehat{\Gamma }_{\gamma \beta
}^\rho -\widehat{\{\rho \alpha \beta \}}\widehat{\Gamma }_{\gamma \delta
}^\rho \right) \eqno(2.10)
$$
(the later is a new object for the\ Riemannian isometry).

The isotopic lifting of the Einstein equations (see the history, details and
references in \cite{Santilli1996}) is written as
$$
\widehat{R}^{\alpha \beta }-\frac 12\widehat{g}^{\alpha \beta }(\widehat{R}+%
\widehat{\Theta })=\widehat{t}^{\alpha \beta }-\widehat{\tau }^{\alpha \beta
},\eqno(2.11)
$$
where $\widehat{t}^{\alpha \beta }$ is a {\bf source isotensor} and $%
\widehat{\tau }^{\alpha \beta }$ is the {\bf stress--energy isotensor} and
there is satisfied the Freud isoidentity \cite{Santilli1996}
$$
\widehat{G}_{~\beta }^\alpha -\frac 12\delta _{~\beta }^\alpha \widehat{%
\Theta }=\widehat{U}_{~\beta }^\alpha +\widehat{\partial }_\rho \widehat{V}%
_{\quad \beta }^{\alpha \rho },\eqno(2.12)
$$
$$
\widehat{U}_{~\beta }^\alpha =-\frac 12\frac{\widehat{\partial }\widehat{%
\Theta }}{\widehat{\partial }\widehat{g}_{\quad \uparrow \alpha }^{\gamma
\delta }}\widehat{g}_{\quad \uparrow \beta }^{\gamma \delta },
$$
$$
\widehat{V}_{\quad \beta }^{\alpha \rho }=\frac 12[\widehat{g}^{\gamma
\delta }\left( \delta _{~\beta }^\alpha \widehat{\Gamma }_{\alpha \delta
}^\rho -\delta _{~\delta }^\alpha \widehat{\Gamma }_{\alpha \beta }^\rho
\right) +
\widehat{g}^{\rho \gamma }\widehat{\Gamma }_{\beta \gamma }^\alpha -\widehat{%
g}^{\alpha \gamma }\widehat{\Gamma }_{\beta \gamma }^\rho +\left( \delta
_{~\beta }^\rho \widehat{g}^{\alpha \gamma }-\delta _{~\beta }^\alpha
\widehat{g}^{\rho \gamma }\right) \widehat{\Gamma }_{\gamma \rho }^\rho ].
$$
Finally, we remark that for antiautomorphic maps of isoduality we have to
modify correspondingly the above presented formulas holding true for
Riemannian isodual spaces $\widehat{{\cal R}}^{(d)}=\widehat{{\cal R}}%
^{(d)}\left( \widehat{x}^{(d)},\widehat{g}^{(d)},\widehat{{\sl R}}%
^{(d)}\right) ,$ over the isodual reals $\widehat{R}^{(d)}=\widehat{R}%
^{(d)}\left( \widehat{n}^{(d)},+,\widehat{\times }^{(d)}\right) $ with
curvature, Ricci, Einstein and so on isodual tensors. For simplicity we
omit such details in this work.

\section{Isobundle Spaces}

This section serves the twofold purpose of establishing of abstract index
denotations and starting the geometric backgrounds of isotopic locally
an\-i\-sot\-rop\-ic extensions of the isoriemannian spaces which are used in
the next sections of the work.

\subsection{Lie--Santilli isoalgebras and isogroups}

The Lie--Santilli isotheory is based on a generalization of the very notion
of numbers and fields. If the Lie's theory is centrally dependent on the
basic $n$--dimensional unit $I=diag\left( 1,1,...,1\right) $ in, for
instance, enveloping algebras, Lie algebras, Lie groups, representation
theory, and so on, the Santilli's main idea  is the
reformulation of the entire conventional theory with respect to the most
general possible, integro--differential isounit. In this subsection we
introduce some necessary definitions and formulas on Lie--Santilli
isoalgebra and isogroups following \cite{Kadeisvili} where details,
developments and basic references on Santilli original result are contained.
A Lie--Santilli algebra is defined as a finite--dimensional isospaces $%
\widehat{L}$ over the isofield $\widehat{F}$ of isoreal or isocomplex
numbers with isotopic element $T$ and isounit $\widehat{I}=T^{-1}.$ In brief
one uses the term isoalgebra (when there is not confusion with isotopies of
non--Lie algebras) which is defined by isolinear isocommutators of type $[A,%
\symbol{94}B]\in \widehat{L}$ satisfying the conditions:%
$$
[A,\symbol{94}B]=-[B,\symbol{94}A],
$$
$$
[A,\symbol{94}[B,\symbol{94}C]]+[B,\symbol{94}[C,\symbol{94}A]]+[C,\symbol{94%
}[A,\symbol{94}B]]=0,
$$
$$
[A*B,\symbol{94}C]=A*[B,\symbol{94}C]+[A,\symbol{94}C]*B
$$
for all $A,B,C\in \widehat{L}.$ The structure functions $\widehat{C}$ of the
Lie--Santilli algebras are introduced according the relations%
$$
[X_i,\symbol{94}X_j]=X_i*X_j-X_j*X_i=
$$
$$
X_iT\left( x,...\right) X_j-X_jT\left( x,...\right) X_i=\widehat{C}%
_{ij}^{\quad k}\left( x,\dot x,\ddot x,...\right) *X_k.
$$
It should be noted that, in fact, the basis $e_k,(k=1,2,...,N)$ of a Lie
algebra $L$ is not changed under isotopy except the renormalization factors $%
\widehat{e}_k:$ the isocommutation rules of the isotopies $\widehat{L}$ are
$$
\left[ \widehat{e}_i,\widehat{e}_j\right] =\widehat{e}_iT\widehat{e}_j-%
\widehat{e}_jT\widehat{e}_i=\widehat{C}_{ij}^{~k}\left( x,\dot x,\ddot
x,...\right) \widehat{e}_k
$$
where $\widehat{C}=CT.$

An isomatrix $\hat M$ is and ordinary matrix whose elements are isoscalars.
 All operations among isomatrices are therefore isotopic.

 The isotrace of a isomatrix $A$ is introduced by using
 the unity $\widehat{I}:$
$$
\widehat{Tr}A=\left( TrA\right) \widehat{I}\in \widehat{F}
$$
where $TrA$ is the usual trace. One holds properties%
$$
\widehat{Tr}(A*B)=\left( \widehat{Tr}A\right) *\left( \widehat{Tr}B\right)
$$
and
$$
\widehat{Tr}A=\widehat{Tr}\left( BAB^{-1}\right) .
$$
The Killing isoform is determined by the isoscalar product%
$$
\left( A,\symbol{94}B\right) =\widehat{Tr}\left[ \left( \widehat{Ad}X\right)
*\left( \widehat{Ad}B\right) \right]
$$
where the isolinear maps are introduced as $\widehat{ad}A\left( B\right) =[A,%
\symbol{94}B],\forall A,B\in \widehat{L}.$ Let $e_k,k=1,2,...,N$ be the
basis of a Lie algebra with an isomorphic map $e_k\rightarrow \widehat{e}_k$
to the basis $\widehat{e}_k$ of a Lie--Santilli isoalgebra $\widehat{L}.$ We
can write the elements in $\widehat{L}$ in local coordinate form. For
instance, considering $A=x^i\widehat{e}_i,B=y^j\widehat{e}_j$ and $C=z^k%
\widehat{e}_k=[A,\symbol{94}B]$ we have%
$$
C=z^k\widehat{e}_k=\left[ A,\symbol{94}B\right] =x^iy^j[\widehat{e}_i,%
\symbol{94}\widehat{e}_j]=x^ix^j\widehat{C}_{ij}^{~k}\widehat{e}_k
$$
and
$$
\left[ \widehat{Ad}A\left( B\right) \right] ^k=\left[ A,\symbol{94}B\right]
^k=x^ix^j\widehat{C}_{ij}^{~k}.
$$
In standard manner there is introduced the {\bf isocartan tensor}
$$
\widehat{q}_{ij}\left( x,\dot x,\ddot x,...\right) =\widehat{C}_{ip}^{~k}%
\widehat{C}_{ik}^{~p}\in \widehat{L}
$$
via the definition
$$
\left( A,\symbol{94}B\right) =\widehat{q}_{ij}x^iy^j.
$$
Considering that $\widehat{L}$ is an isoalgebra with generators $X_k$ and
isounit $\widehat{I}=T^{-1}>0$ the isodual Lie--Santilli algebras $\widehat{L%
}^d$ of $\widehat{L}$ (we note that $\widehat{L}$ and $\widehat{L}^d$ are
(anti) isomorphic).

The conventional structure of the Lie theory admits a conventional isotopic
lifting. Let give some examples. The general isolinear and isocomplex
Lie--Santilli algebras $\widehat{gl}\left( n,\widehat{C}\right) $ are
introduced as the vector isospaces of all $n\times n$ isocomplex matrices over $%
\widehat{C}.$ For the isoreal numbers $\widehat{R}$ we shall write $\widehat{%
gl}\left( n,\widehat{R}\right) .$ By using ''hats'' we denote respectively
the special, isocomplex, isolinear isoalgebra $\widehat{sl}\left( n,\widehat{C}%
\right) $ and the isoorthogonal algebra $\widehat{o}\left( n\right) .$

A right Lie--Santilli isogroup $\widehat{G}r$ on an isospace $\widehat{S}%
\left( x,\widehat{F}\right) $ over an isofield $\widehat{F},\widehat{I}%
=T^{-1}$ (in brief isotransformation group or isogroup) is introduced in
standard form but with respect to isonumbers and isofields as a group which
maps each element $x\in \widehat{S}\left( x,\widehat{F}\right) $ into a new
element $x^{\prime }\in \widehat{S}\left( x,\widehat{F}\right) $ via the
isotransformations $x^{\prime }=\widehat{U}*x=\widehat{U}Tx,$ where $T$ is
fixed such that

1. The map $\left( U,x\right) \rightarrow \widehat{U}*x$ of $\widehat{G}%
r\times \widehat{S}\left( x,\widehat{F}\right) $ onto $\widehat{S}\left( x,%
\widehat{F}\right) $ is isodifferentiable;

2. $\widehat{I}*\widehat{U}=\widehat{U}*\widehat{I}=\widehat{U},~\forall
\widehat{U}\in \widehat{G}r;$

3. $\widehat{U}_1*(\widehat{U}_2*x)=\left( \widehat{U}_1*\widehat{U}%
_2\right) *x,~\forall x\in \widehat{S}\left( x,\widehat{F}\right) $ and $%
\widehat{U}_1,\widehat{U}_2\in \widehat{G}r.$

We can define accordingly a left isotransformation group.

\subsection{Fiber isobundles}

Prof. Santilli identified the foundations of the isosympletic geometry in
 the work \cite{Santilli1996}. In this section we present, apparently for
 the first time, the isotopies of fibre bundles and related topics.

The notion of locally trivial fiber isobundle naturally generalizes that of
the isomanifold. The fiber isobundles will be used to get some results in
isogeometry as well as to build geometrical models for physical isotheories.
In general the proofs, being corresponding reformulation in isotopic manner
of standard results, will be omitted. The reader is referred to some
well--known books containing the theory of fibre bundles and the
mathematical foundations of the Lie--Santilli isotheory.

Let $\widehat{G}r$ be a Lie--Santilli isogroup which acts isodifferentiably
and effectively on a isomanifold $\widehat{V},$ i.e. every element $%
\widehat{q}\in \widehat{G}r$ defines an isotopic diffeomorphism $L_{\widehat{%
G}r}:\widehat{V}\rightarrow \widehat{V}.$

As a rule, all isomanifolds are assumed to be isocontinuous, finite
dimensional and having the isotopic variants of the conditions to be
Hausdorff, paracompact and isoconnected; all isomaps are isocontinous.

A locally trivial {\bf fibre isobundle} is defined by the data\\ $\left(
\widehat{E},\widehat{p},\widehat{M},\widehat{V},\widehat{G}r\right) ,$ where
$\widehat{M}$ (the base isospace) and $\widehat{E}$ (the total isospace) are
isomanifolds, $\widehat{E},\widehat{p}:\widehat{E}\rightarrow \widehat{M}$
is a surjective isomap and the following conditions are satisfied:

1/ the isomanifold $\widehat{M}$ can be covered by a set ${\cal E}$ of open
isotopic sets $\widehat{U},\widehat{W},...$ such that for every open set $%
\widehat{U}$ there exist a bijective isomap $\widehat{\varphi }_{\widehat{U}%
}:\widehat{p}^{-1}\left( \widehat{U}\right) \rightarrow \widehat{U}\times
\widehat{V}$ so that $\widehat{p}\left( \varphi _{\widehat{U}}^{-1}\left(
\widehat{x},\widehat{y}\right) \right) =\widehat{x},~\forall \widehat{x}\in
\widehat{U},\forall \widehat{y}\in \widehat{V};$

2/ if $\widehat{x}\in \widehat{U}\cap \widehat{W}\neq \oslash ,$ than $%
\widehat{\varphi }_{\widehat{W},x}\circ \widehat{\varphi }_{\widehat{U}%
,x}^{-1}:\widehat{V}$ $\rightarrow \widehat{V}$ is an isotopic
diffeomorphism $L_{\widehat{g}r}$ with $\widehat{g}r\in \widehat{G}r$ where $%
\widehat{\varphi }_{\widehat{U},x}$ denotes the restriction of $\widehat{%
\varphi }_{\widehat{U}}$ to $p^{-1}\left( \widehat{x}\right) $ and $\widehat{%
U},\widehat{W}\in {\cal E;}$

3/ the isomap $q_{\widehat{U}\widehat{V}}:\widehat{U}\cap \widehat{V}%
\rightarrow \widehat{G}r$ defined by structural isofunctions $q_{\widehat{U}%
\widehat{V}}\left( \widehat{x}\right) =\widehat{\varphi }_{\widehat{W}%
,x}\circ \widehat{\varphi }_{\widehat{U},x}^{-1}$ is isocontinous.

Let $In_U$ and $In_V$ be sets of indices and denote by $\left( \widehat{U}%
_\alpha ,\widehat{\varphi }_\alpha \right) _{\alpha \in In_U}$ and $\left(
\widehat{V}_\beta ,\widehat{\psi }_\beta \right) _{\beta \in In_V}$ be
correspondingly isocontinous atlases on $\widehat{U}_\alpha $ and $\widehat{V%
}_\beta .\,$ One obtains an isotopic topology on $\widehat{E}$ for which the
bijections $\widehat{\varphi }_{\widehat{U}},\widehat{\varphi }_{\widehat{W}%
,x}...$ become isotopic homeomorphisms. Denoting respectively by $n$ and $m$
the dimensions of the isomanifolds $\widehat{M}$ and $\widehat{V}$ we can
define the isotopic maps
$$
\phi _{\alpha \beta }:\varpi _{\alpha \beta }\rightarrow \widehat{R}%
^{n+m},\phi _{\alpha \beta }=\left( \widehat{\varphi }_\alpha \times
\widehat{\psi }_\beta \right) \circ \varphi _{\widehat{U}}^{\alpha \beta }
$$
where $\varphi _{\widehat{U}}^{\alpha \beta }$ is the restriction to the
isomap $\widehat{\varphi }_{\widehat{U}}$ to $\varpi _{\alpha \beta }.$ Than
the set\\ $\left( \varpi _{\alpha \beta },\phi _{\alpha \beta }\right)
_{\left( \alpha ,\beta \right) \in In_U\times In_V}$ is a isocontinous atlas
on $\widehat{E}.$

A locally trivial {\bf principal isobundle} $\left( \widehat{P},\widehat{\pi
},\widehat{M},\widehat{G}r\right) $ is a fibre isobundle $\left( \widehat{E},%
\widehat{p},\widehat{M},\widehat{V},\widehat{G}r\right) $ for which the type
fibre coincides with the structural group, $\widehat{V}=\widehat{G}r$ and the
action of $\widehat{G}r$ on $\widehat{G}r$ is given by the left isotransform
$\widehat{L}_q\left( a\right) =qa,\forall q,a\in \widehat{G}r.$

The structural functions of the principal isobundle $\left( \widehat{P},%
\widehat{\pi },\widehat{M},\widehat{G}r\right) $ are
$$
q_{\widehat{U}\widehat{W}}:\widehat{U},\widehat{W}\rightarrow \widehat{G}%
r,~q_{\widehat{U}\widehat{W}}\left( \widehat{\pi }\left( u\right) \right) =%
\widehat{\varphi }_{\widehat{W}}\left( u\right) \circ \widehat{\varphi }_{%
\widehat{U}}^{-1}\left( u\right) ,u\in \widehat{\pi }^{-1}\left( \widehat{U}%
\cap \widehat{W}\right) .
$$

A {\bf morphism of principal isobundles} $\left( \widehat{P},\widehat{\pi },%
\widehat{M},\widehat{G}r\right) $ and \\ $\left( \widehat{P}^{\prime },%
\widehat{\pi }^{\prime },\widehat{M}^{\prime },\widehat{G}r^{\prime }\right)
$ is a pair $\left( \widehat{f},\widehat{f}^{\prime }\right) $ of isomaps
for which the following conditions hold:

1/ $\widehat{f}:\widehat{P}\rightarrow \widehat{P}^{\prime }$ is a
isocontinous isomap,

2/ $\widehat{f}:\widehat{G}r\rightarrow \widehat{G}r^{\prime }$ is an
isotopic morphism of Lie--Santilli isogroups.

3/ $\widehat{f}\left( \widehat{u}\widehat{q}\right) =f\left( \widehat{u}%
\right) f^{\prime }\left( \widehat{q}\right) ,\widehat{u}\in \widehat{P},%
\widehat{q}\in \widehat{G}r.$

We can define isotopic isomorphisms, automorphisms and subbundles in a usual
manner but with respect to isonumbers, isofields, isogroups and isomanifold
when corresponding isotopic transforms and maps provide the isotopic
properties.

A {\bf isotopic subbundle} $\left( \widehat{P},\widehat{\pi },\widehat{M},%
\widehat{G}r\right) $ of the principal isobundle\\ $\left( \widehat{P}%
^{\prime },\widehat{\pi }^{\prime },\widehat{M}^{\prime },\widehat{G}%
r^{\prime }\right) $ is called a reduction of the structural isogroup $%
\widehat{G}r^{\prime }$ to $\widehat{G}r.$

An isotopic frame (isoframe) in a point $\widehat{x}\in \widehat{M}$ is a
set of $n$ linearly independent isovectors tangent to $\widehat{M}$ in $%
\widehat{x}.$ The set $\widehat{L}\left( \widehat{M}\right) $ of all
isoframes in all points of $\widehat{M}$ can be naturally provided (as in
the non isotopic case, see, for instance, \cite{KobayashiNomizu}) with an
isomanifold structure. The principal isobundle $\left( \widehat{L}\left(
\widehat{M}\right) ,\widehat{\pi },\widehat{M},\widehat{Gl}\left( n,\widehat{%
R}\right) \right) $of isoframes on $\widehat{M},$ denoted in brief also by $%
\widehat{L}\left( \widehat{M}\right) ,$ has $\widehat{L}\left( \widehat{M}%
\right) $ as the total space and the general linear isogroup\\ $\widehat{Gl}%
\left( n,\widehat{R}\right) $ as the structural isogroup.

Having introduced the isobundle $\widehat{L}\left( \widehat{M}\right) $ we
can give define an {\bf isotopic }$G${\bf --structure} on a isomanifold $%
\widehat{M}$ is a subbundle $\left( \widehat{P},\widehat{\pi },\widehat{M},%
\widehat{G}r\right) $ of the principal isobundle $\widehat{L}\left( \widehat{%
M}\right) $ being an isotopic reduction of the structural isogroup $\widehat{%
Gl}\left( n,\widehat{R}\right) $ to a isotopic subgroup $\widehat{G}r$ of it.

A very important class of bundle spaces used for modeling of locally
anisotropic interactions is that of vector bundles. We present here the
necessary isotopic generalizations.

An locally trivial {\bf iso\-vec\-tor bund\-le} (equi\-valently, {\bf vector
iso\-bundle, v--isobundle}) $\left( \widehat{E},\widehat{p},\widehat{M},%
\widehat{V},\widehat{G}r\right) $ is defined as a corresponding fibre
isobundle if $\widehat{V}$ is a linear iso\-spa\-ce and $\widehat{G}r$ is
the Lie--Santilli isogroup of iso\-top\-ic authomorphisms of $\widehat{V}.$

For $\widehat{V}=\widehat{R}^m$ and $\widehat{G}r=\widehat{Gl}\left( m,%
\widehat{R}\right) $ the v--isobundle\\ $\left( \widehat{E},\widehat{p},%
\widehat{M},\widehat{R}^m,\widehat{Gl}\left( m,\widehat{R}\right) \right) $
is denoted shortly as $\widehat{\xi }=\left( \widehat{E},\widehat{p},%
\widehat{M}\right) .$ Here we also note that the transformations of the
isocoordinates $\left( \widehat{x}^k,\widehat{y}^a\right) \rightarrow \left(
\widehat{x}^{k^{\prime }},\widehat{y}^{a^{\prime }}\right) $ on $\widehat{%
\xi }$ are of the form
$$
\widehat{x}^{k^{\prime }}=\widehat{x}^{k^{\prime }}\left( \widehat{x}^1,...,%
\widehat{x}^n\right) ,\quad rank\left( \frac{\widehat{\partial }\widehat{x}%
^{k^{\prime }}}{\widehat{\partial }\widehat{x}^k}\right) =n
$$
$$
\widehat{y}^{a^{\prime }}=\widehat{Y}_a^{a^{\prime }}\left( x\right)
y^a,\quad \widehat{Y}_a^{a^{\prime }}\left( x\right) \in \widehat{Gl}\left(
m,\widehat{R}\right) .
$$

A local isocoordinate parametrization of $\widehat{\xi }$ naturally defines
an isocoordinate basis%
$$
\frac \partial {\partial \widehat{u}^\alpha }=\left( \frac \partial
{\partial x^i},\frac \partial {\partial y^a}\right) ,\eqno(3.1)
$$
in brief we shall write $\widehat{\partial }_\alpha =(\widehat{\partial }_i,%
\widehat{\partial }_a),$ and the reciprocal to (3.1) coordinate basis
$$
d\widehat{u}^\alpha =(d\widehat{x}^i,d\widehat{y}^a),
$$
or, in brief, $\widehat{d}^\alpha =(\widehat{d}^i,\widehat{d}^a),$ which is
uniquely defined from the equations%
$$
\widehat{d}^\alpha \circ \widehat{\partial }_\beta =\delta _\beta ^\alpha ,
$$
where $\delta _\beta ^\alpha $ is the Kronecher symbol and by $"\circ $$"$
we denote the inner (scalar) product in the isotangent isobundle $\widehat{%
T\xi }$ (see the definition of isodifferentials and partial isoderivations
in (2.1)--(2.3)). Here we note that the tangent isobundle (in brief
t--isobundle) of a isomanifold $\widehat{M},$ denoted as $\widehat{TM}=\cup
_{x\in \widehat{M}}\widehat{T_xM},$ where $\widehat{T_xM}$ is tangent
isospaces of tangent isovectors in the point $\widehat{x}\in $ $\widehat{M},$
is defined as a v--isobundle $\widehat{E}=\widehat{TM}.$ By $\widehat{TM}%
^{*} $ we define the dual (not confusing with isotopic dual) of the
t--isobundle $\widehat{TM}.$ We note that for $\widehat{TM}$ and $\widehat{TM%
}^{*}$ isobundles the fibre and the base have both the same dimension and it
is not necessary to distinguish always the fiber and base indices by
different letters.

\section{ Nonlinear and Distinguished Isocon\-nec\-ti\-ons}

The concept of {\bf nonlinear connection,} in brief, N--connection, is
fundamental in the geometry of locally anisotropic spaces (in brief,
 la--spaces,
see a detailed study and basic references in \cite{MironAnastasiei}). Here
 it should be noted that we consider the term la--space
in a more general context than
 G. Yu. Bogoslovsky \cite{Bogoslovsky} which uses it for a class of Finsler
spaces and Finsler gravitational theories. In our papers
\cite{VacaruGoncharenko,VacaruJMP,VacaruAP,VacaruSupergravity,
VacaruStochastics,VacaruCL} the la--spaces and la--superspaces are
 respectively modelled on vector bundles and vector superbundles
 enabled with compatible nonlinear and distinguished connections
 and metric structures (as particular cases, on tangent bundles and
 superbundles, for corresponding classes of metrics and nonlinear
 connections, one constructs  generalized Lagrange and Finsler spaces and
 superspaces). The geometrical objects on a la--space are called
{\bf distinguished} (see detailes in \cite{MironAnastasiei}) if they are
 compatible with the N--connection structure (one considers,
 for instance, distinguished connections and distinguished tensors,
 in brief, d--connections and d--tensors).

 In this section we study, apparently for the first time, the isogeometry
 of N--connection in vector isobundle. We note that a type of
 generic nonlinearity is contained by definition in the  structure
 of isospace (it can be associated to a corresponding class of nonlinear
 isoconnections which can be turned into linear ones under corresponding
 isotopic transforms). As to a general N--connection introduced as
 a  global decompositions of a vector isobundle into  horizontal and vertical
  isotopic subbundles (see below) it can not be isolinearized if its
 isocurvature is nonzero.

Let consider a v--isobundle $\widehat{\xi }=\left( \widehat{E},\widehat{p},%
\widehat{M}\right) $ whose type fibre is $\widehat{R}^m$ and $\widehat{p}^T:%
\widehat{TE}\rightarrow \widehat{TM}$ is the isodifferential of the isomap $%
\widehat{p}.$ The kernel of the isomap $\widehat{p}^T$ (which is a
fibre--preserving isotopic morphism of the t--isobundle $\left( \widehat{TE},%
\widehat{\tau }_E,\widehat{E}\right) $ to $\widehat{E}$ and of t--isobundle $%
\left( \widehat{TM},\widehat{\tau }_M,\widehat{M}\right) $ to $\widehat{M})$
defines the vertical isotopic subbundle $\left( \widehat{VE},\widehat{\tau }%
_V,\widehat{E}\right) $ over $\widehat{E}$ being an isovector subbundle of
the v--isobundle $\left( \widehat{TE},\widehat{\tau }_E,\widehat{E}\right) .$

An isovector $\widehat{X}_u$ tangent to $\widehat{E}$ in a point $\widehat{u}%
\in \widehat{E}$ locally defined by the decomposition $\widehat{X}^i\widehat{%
\partial }_i+\widehat{Y}^a\widehat{\partial }_a$ is locally represented by
the isocoordinates%
$$
\widehat{X}_u=\left( \widehat{x},\widehat{y},\widehat{X},\widehat{Y}\right)
=\left( \widehat{x}^i,\widehat{y}^a,\widehat{X}^i,\widehat{Y}^a\right) .
$$
Since $\widehat{p}^T\left( \widehat{\partial }_a\right) =0$ it results that $%
\widehat{p}^T\left( \widehat{x},\widehat{y},\widehat{X},\widehat{Y}\right)
=\left( \widehat{x},\widehat{X}\right) .$ We also consider the isotopic
imbedding map $\widehat{i}:\widehat{VE}\rightarrow \widehat{TE}$ and the
isobundle of inverse image $\widehat{p}^{*}\widehat{TM}$ of the $\widehat{p}:%
\widehat{E}\rightarrow \widehat{M}$ and define in result the isomap $%
\widehat{p}!:\widehat{TE}\rightarrow $ $\widehat{p}^{*}\widehat{TM},\widehat{%
p}!\left( \widehat{X}_u\right) =\left( \widehat{u},\widehat{p}^T\left(
\widehat{X}_u\right) \right) $ for which one holds $Ker~\widehat{p}!=Ker~%
\widehat{p}^T=\widehat{VE}.$

A {\bf nonlinear isoconnection,} (in brief, {\bf N--isoconnection)}
in the iso\-vec\-t\-or bundle $\widehat{\xi }=\left(
\widehat{E},\widehat{p},\widehat{M}\right) $ is defined as a splitting on
the left of the exact sequence of isotopic maps%
$$
0\longrightarrow \widehat{VE}\stackrel{\widehat{i}}{\longrightarrow }%
\widehat{TE}\stackrel{\widehat{p}!}{\longrightarrow }\widehat{TE}/\widehat{VE%
}\longrightarrow 0
$$
that is an isotopic morphism of vector isobundles $\widehat{C}:\widehat{TE}%
\rightarrow \widehat{VE}$ such that $\widehat{C}\circ \widehat{i}$ is the
identity on $\widehat{VE}.$

The kernel of the isotopic morphism $\widehat{C}$ is a isovector subbundle
of $\left( \widehat{TE},\widehat{\tau }_E,\widehat{E}\right) $ and will be
called the horizontal isotopic subbundle\\ $\left( \widehat{HE},\widehat{%
\tau }_H,\widehat{E}\right) .$

As a consequence of the above presented definition
 we can consider that a N--isoconnection in
v--isobundle $\widehat{E}$ is a isotopic distribution $\{\widehat{N}:%
\widehat{E}_u\rightarrow H_u\widehat{E},T_u\widehat{E}=H_u\widehat{E}\oplus
V_u\widehat{E}\}$ on $\widehat{E}$ such that it is defined a global
decomposition, as a Whitney sum, into horizontal,$\widehat{HE},$ and
vertical, $\widehat{VE},$ subbundles of the tangent isobundle $\widehat{TE}:$
$$
\widehat{TE}=\widehat{HE}\oplus \widehat{VE}.\eqno(4.1)
$$

Locally a N--isoconnection in $\widehat{\xi }$ is given by its components $%
\widehat{N}_i^a(\widehat{u}) = \widehat{N}_i^a(\widehat{x},\widehat{y})$ $=
N_{\widehat{i}}^{\widehat{a}}(\widehat{x},\widehat{y})$ with respect to
local isocoordinate bases (2.14) and (2.15)):
$$
\widehat{{\bf N}}=\widehat{N}_i^a(\widehat{u})\widehat{d}^i\otimes \widehat{%
\partial }_a.
$$

We note that a linear isoconnection in a v--isobundle $\widehat{\xi }$ can be
considered as a particular case of a N--isoconnection when $\widehat{N}_i^a(%
\widehat{x},\widehat{y})=\widehat{K}_{bi}^a\left( \widehat{x}\right)
\widehat{y}^b,$ where isofunctions $\widehat{K}_{ai}^b\left( \widehat{x}%
\right) $ on the base $\widehat{M}$ are called the isochristoffel
coefficients.

To coordinate locally geometric constructions with the global splitting of
isobundle defined by a N--isoconnection structure, we have to introduce a
locally adapted isobasis (la---isobasis, la---isoframe):
$$
\frac{\widehat{\delta }}{\delta \widehat{u}^\alpha }=\left( \frac{\widehat{%
\delta }}{\delta \widehat{x}^i}=\widehat{\partial }_i-\widehat{N}_i^a\left(
\widehat{u}\right) \widehat{\partial }_a,\frac{\widehat{\partial }}{\partial
\widehat{y}^a}\right) ,\eqno(4.2)
$$
or, in brief, $\widehat{\delta }_\alpha =\delta _{\widehat{\alpha }}=\left(
\widehat{\delta }_i,\widehat{\partial }_a\right) ,$ and its dual la-isobasis%
$$
\widehat{\delta }\widehat{u}^\alpha =\left( \widehat{\delta }\widehat{x}^i=%
\widehat{d}\widehat{x}^i,\widehat{\delta }\widehat{y}^a+\widehat{N}%
_i^a\left( \widehat{u}\right) \widehat{d}\widehat{x}^i\right) ,\eqno(4.3)
$$
or, in brief,{\bf \ }$\ \widehat{\delta }\ ^\alpha =$ $\left( \widehat{d}^i,%
\widehat{\delta }^a\right) .$ We note that isooperators (4.2) and (4.3)
generalize correspondingly the partial isoderivations and isodifferentials
(2.1)--(2.3) for the case when a N--isoconnection is defined.

The{\bf \ nonholonomic isocoefficients}$\widehat{{\bf \ w}}=\{\widehat{w}%
_{\beta \gamma }^\alpha \left( \widehat{u}\right) \}$ of la--isoframes are
defined as%
$$
\left[ \widehat{\delta }_\alpha ,\widehat{\delta }_\beta \right] =\widehat{%
\delta }_\alpha \widehat{\delta }_\beta -\widehat{\delta }_\beta \widehat{%
\delta }_\alpha =\widehat{w}_{\beta \gamma }^\alpha \left( \widehat{u}%
\right) \widehat{\delta }_\alpha .
$$

The {\bf algebra of tensorial distinguished isofields} $D\widehat{T}\left(
\widehat{\xi }\right) $ (d--iso\-fi\-elds, d--isotensors, d--tensor isofield,
d--isobjects) on $\widehat{\xi }$ is introduced as the tensor algebra ${\cal %
T}=\{\widehat{{\cal T}}_{qs}^{pr}\}$ of the v--isobundle $\widehat{\xi }_{\left(
d\right) },$ $\widehat{p}_d:\widehat{HE}\oplus \widehat{VE}\rightarrow
\widehat{E}.$ An element $\widehat{{\bf t}}\in \widehat{{\cal T}}_{qs}^{pr},$
d--tensor isofield of type $\left(
\begin{array}{cc}
p & r \\
q & s
\end{array}
\right) ,$ can be written in local form as%
$$
\widehat{{\bf t}}=\widehat{t}_{j_1...j_qb_1...b_r}^{i_1...i_pa_1...a_r}%
\left( u\right) \widehat{\delta }_{i_1}\otimes ...\otimes \widehat{\delta }%
_{i_p}\otimes \widehat{\partial }_{a_1}\otimes ...\otimes \widehat{\partial }%
_{a_r}\otimes
$$
$$
\widehat{d}^{j_1}\otimes ...\otimes \widehat{d}^{j_q}\otimes \widehat{\delta
}^{b_1}...\otimes \widehat{\delta }^{b_r}.
$$

We shall respectively use denotations ${\cal X}\left( \widehat{\xi }\right) $
(or ${\cal X}{\left( \widehat{M}\right) ),\ }\Lambda ^p\left( \widehat{\xi }%
\right) $\\ (or $\Lambda ^p\left( \widehat{M}\right) )$ and ${\cal F}\left(
\widehat{\xi }\right) $ (or ${\cal F}\left( \widehat{M}\right) $) for the
isotopic module of d-vector isofields on $\widehat{\xi }$ (or $\widehat{M}$
), the exterior algebra of p-forms on $\widehat{\xi }$ (or $\widehat{M}$ )
and the set of real functions on $\widehat{\xi }$ (or $\widehat{M}$ ).

In general, d--objects on $\widehat{\xi }$ are introduced as geometric
objects with various isogroup and isocoordinate transforms coordinated with
the N--connection isostructure on $\widehat{\xi }.$ For example, a
d--connection $\widehat{D}$ on $\widehat{\xi } $ is defined as a isolinear
connection $\widehat{D}$ on $\widehat{E}$ conserving under a parallelism the
global decomposition (4.1) into horizontal and vertical subbundles of $%
\widehat{TE}.$

A N--connection in $\widehat{\xi }$ induces a corresponding decomposition of
d--iso\-ten\-sors into sums of horizontal and vertical parts, for example,
for every d--isovector $\widehat{X}\in {\cal X}\left( \widehat{\xi }\right) $
and 1--form $\widetilde{X}\in \Lambda ^1\left( \widehat{\xi }\right) $ we
have respectively
$$
\widehat{X}=\widehat{hX}+\widehat{vX}{\bf \ \quad }\mbox{and \quad }%
\widetilde{X}=h\widetilde{X}+v\widetilde{X}~.
$$
In consequence, we can associate to every d--covariant isoderivation $%
\widehat{D_X}$ $=\widehat{X}\circ \widehat{D}$ two new operators of h- and
v-covariant isoderivations defined respectively as
$$
\widehat{D}_X^{(h)}\widehat{Y}=\widehat{D}_{hX}\widehat{Y}\quad
\mbox{ and \quad }\widehat{D}_X^{\left( v\right) }\widehat{Y}=\widehat{D}%
_{vX}\widehat{Y}{\bf ,\quad }\forall \widehat{Y}{\bf \in }{\cal X}\left(
\widehat{\xi }\right)
$$
for which the following conditions hold:%
$$
\widehat{D}_X\widehat{Y}{\bf =}\widehat{D}_X^{(h)}\widehat{Y}{\bf \ +}%
\widehat{D}_X^{(v)}\widehat{Y}{\bf ,}\eqno(4.4)
$$
$$
\widehat{D}_X^{(h)}f=(h\widehat{X}{\bf )}f\mbox{ \quad and\quad }\widehat{D}%
_X^{(v)}f=(v\widehat{X}{\bf )}f,\quad \widehat{X},\widehat{Y}{\bf \in }{\cal %
X}\left( \widehat{\xi }\right) ,f\in {\cal F}\left( \widehat{M}\right) .
$$

An {\bf isometric structure }
$\widehat{{\bf G}}$ in the total space $\widehat{E} $
of v--isobundle\\ $\widehat{\xi } = $ $\left( \widehat{E},\widehat{p},\widehat{%
M}\right) $ over a connected and paracompact base $\widehat{M}$ is
introduced as a symmetrical covariant tensor isofield of type $\left(
0,2\right) $, $\widehat{G}_{\alpha \beta ,}$ being nondegenerate and of
constant signature on $\widehat{E}.$

Nonlinear isoconnection $\widehat{{\bf N}}$ and isometric $\widehat{{\bf G}}$
structures on $\widehat{\xi }$ are mutually compatible it there are
satisfied the conditions:
$$
\widehat{{\bf G}}\left( \widehat{\delta }_i,\widehat{\partial }_a\right) =0
$$
which in component form are written as
$$
\widehat{G}_{ia}\left( \widehat{u}\right) -\widehat{N}_i^b\left( \widehat{u}%
\right) \widehat{h}_{ab}\left( \widehat{u}\right) =0,\eqno(4.5)
$$
where $\widehat{h}_{ab}=\widehat{{\bf G}}\left( \widehat{\partial }_a,%
\widehat{\partial }_b\right) $ and $\widehat{G}_{ia}=\widehat{{\bf G}}\left(
\widehat{\partial }_i,\widehat{\partial }_a\right) $ (the matrix $\widehat{h}%
^{ab}$ is inverse to $\widehat{h}_{ab}).$

In consequence one obtains the following decomposition of isotopic metric :%
$$
\widehat{{\bf G}}(\widehat{X},\widehat{Y}){\bf =}\widehat{{\bf hG}}(\widehat{%
X},\widehat{Y})+\widehat{{\bf vG}}(\widehat{X},\widehat{Y})
$$
where the d-tensor $\widehat{{\bf hG}}(\widehat{X},\widehat{Y}){\bf =}%
\widehat{{\bf G}}(\widehat{hX},\widehat{hY})$ is of type $\left(
\begin{array}{cc}
0 & 0 \\
2 & 0
\end{array}
\right) $ and the d--isotensor
 $\widehat{{\bf vG}}(\widehat{X},\widehat{Y}){\bf =%
}\widehat{{\bf G}}(v\widehat{X},v\widehat{Y})$ is of type $\left(
\begin{array}{cc}
0 & 0 \\
0 & 2
\end{array}
\right) .$ With respect to la--isobasis (4.3) the d--isometric  is written
as%
$$
\widehat{{\bf G}}=\widehat{g}_{\alpha \beta }\left( \widehat{u}\right)
\widehat{\delta }^\alpha \otimes \widehat{\delta }^\beta =\widehat{g}%
_{ij}\left( \widehat{u}\right) \widehat{d}^i\otimes \widehat{d}^j+\widehat{h}%
_{ab}\left( \widehat{u}\right) \widehat{\delta }^a\otimes \widehat{\delta }%
^b,\eqno(4.6)
$$
where $\widehat{g}_{ij}=\widehat{{\bf G}}\left( \widehat{\delta }_i,\widehat{%
\delta }_j\right) .$

A metric isostructure of type (4.6) on $\widehat{E}$ with components
satisfying constraints (4.3)) defines an adapted to the given N--isoconnection
inner (d--scalar) isoproduct on the tangent isobundle $\widehat{TE}.$

A d--isoconnection $\widehat{D}_X$ is {\bf compatible } with an isometric $%
\widehat{{\bf G}}$ on $\widehat{\xi }$ if%
$$
\widehat{D}_X\widehat{{\bf G}}=0,\forall \widehat{X}{\bf \in }\widehat{{\cal %
X}}\left( \widehat{\xi }\right) .
$$
Locally adapted components $\widehat{\Gamma }_{\beta \gamma }^\alpha $ of a
d--isoconnection $\widehat{D}_\alpha =(\widehat{\delta }_\alpha \circ \widehat{D}%
)$ are defined by the equations%
$$
\widehat{D}_\alpha \widehat{\delta }_\beta =\widehat{\Gamma }_{\alpha \beta
}^\gamma \widehat{\delta }_\gamma ,
$$
from which one immediately follows%
$$
\widehat{\Gamma }_{\alpha \beta }^\gamma \left( \widehat{u}\right) =\left(
\widehat{D}_\alpha \widehat{\delta }_\beta \right) \circ \widehat{\delta }%
^\gamma .\eqno(4.7)
$$
The operations of h- and v--covariant
 isoderivations, $\widehat{D}_k^{(h)}=\{%
\widehat{L}_{jk}^i,\widehat{L}_{bk\;}^a\}$ and $\widehat{D}_c^{(v)}=\{%
\widehat{C}_{jk}^i,\widehat{C}_{bc}^a\}$ (see (4.4)), are introduced as
corresponding h-- and v--parametrizations of (4.7):%
$$
\widehat{L}_{jk}^i=\left( \widehat{D}_k\widehat{\delta }_j\right) \circ
\widehat{d}^i,\quad \widehat{L}_{bk}^a=\left( \widehat{D}_k\widehat{\partial
}_b\right) \circ \widehat{\delta }^a\eqno(4.8)
$$
and%
$$
\widehat{C}_{jc}^i=\left( \widehat{D}_c\widehat{\delta }_j\right) \circ
\widehat{d}^i,\quad \widehat{C}_{bc}^a=\left( \widehat{D}_c\widehat{\partial
}_b\right) \circ \widehat{\delta }^a.\eqno(4.9)
$$
Components (4.8) and (4.9), $\widehat{D}\widehat{\Gamma }=\left( \widehat{L}%
_{jk}^i,\widehat{L}_{bk}^a,\widehat{C}_{jc}^i,\widehat{C}_{bc}^a\right) ,$
completely defines the local action of a d--isoconnection $\widehat{D}$ in $%
\widehat{\xi }.$ For instance, taken a d--tensor isofield of type $\left(
\begin{array}{cc}
1 & 1 \\
1 & 1
\end{array}
\right) ,$
$$
\widehat{{\bf t}}=\widehat{t}_{jb}^{ia}\widehat{\delta }_i\otimes \widehat{%
\partial }_a\otimes \widehat{\partial }^j\otimes \widehat{\delta }^b,
$$
and a d-vector $\widehat{{\bf X}}=\widehat{X}^i\widehat{\delta }_i+\widehat{X%
}^a\widehat{\partial }_a$ we have%
$$
\widehat{D}_X\widehat{{\bf t}}{\bf =}\widehat{D}_X^{(h)}\widehat{{\bf t}}%
{\bf +}\widehat{D}_X^{(v)}\widehat{{\bf t}}{\bf =}\left( \widehat{X}^k%
\widehat{t}_{jb|k}^{ia}+\widehat{X}^c\widehat{t}_{jb\perp c}^{ia}\right)
\widehat{\delta }_i\otimes \widehat{\partial }_a\otimes \widehat{d}^j\otimes
\widehat{\delta }^b,
$$
where the{\bf \ h--covariant and v--covariant isoderivatives} are written
respectively as%
$$
\widehat{t}_{jb|k}^{ia}=\frac{\widehat{\delta }\widehat{t}_{jb}^{ia}}{\delta
\widehat{x}^k}+\widehat{L}_{hk}^i\widehat{t}_{jb}^{ha}+\widehat{L}_{ck}^a%
\widehat{t}_{jb}^{ic}-\widehat{L}_{jk}^h\widehat{t}_{hb}^{ia}-\widehat{L}%
_{bk}^c\widehat{t}_{jc}^{ia}
$$
and
$$
\widehat{t}_{jb\perp c}^{ia}=\frac{\widehat{\partial }\widehat{t}_{jb}^{ia}}{%
\partial \widehat{y}^c}+\widehat{C}_{hc}^i\widehat{t}_{jb}^{ha}+\widehat{C}%
_{dc}^a\widehat{t}_{jb}^{id}-\widehat{C}_{jc}^h\widehat{t}_{hb}^{ia}-%
\widehat{C}_{bc}^d\widehat{t}_{jd}^{ia}.
$$
For a scalar isofunction $f\in {\cal F(}\widehat{\xi })$ we have
$$
\widehat{D}_k^{(h)}=\frac{\widehat{\delta }f}{\delta \widehat{x}^k}=\frac{%
\widehat{\partial }f}{\partial \widehat{x}^k}-\widehat{N}_k^a\frac{\widehat{%
\partial }f}{\partial \widehat{y}^a}\mbox{ and }\widehat{D}_c^{(v)}f=\frac{%
\widehat{\partial }f}{\partial \widehat{y}^c}.
$$
We emphasize that the geometry of connections in a v--isobundle $\widehat{%
\xi }$ is very reach. For instance, if a triple of fundamental isogeometric
objects $(\widehat{N}_i^a\left( \widehat{u}\right) ,$ $\widehat{\Gamma }%
_{\beta \gamma }^\alpha \left( \widehat{u}\right) ,$ $\widehat{G}_{\alpha
\beta }\left( \widehat{u}\right) )$ is fixed on $\widehat{\xi },$ a
multi--isoconnection structure (with corresponding rules of covariant
isoderivation, which are, or not, mutually compatible and with the same, or
not, induced d--scalar products in $\widehat{TE})$ is defined.

Let enumerate some of isoconnections and covariant isoderivations which can
present interest in investigation of locally anisotropic and
 homogeneous gravitational and
matter field isotopic interactions:

\begin{enumerate}
\item  Every N--isoconnection in $\widehat{\xi }$ with coefficients $\widehat{N}%
_i^a\left( \widehat{x},\widehat{y}\right) $ being isodifferentiable on
y-variables induces a structure of isolinear isoconnection $\widetilde{N}%
_{\beta \gamma }^\alpha ,$ where $\widetilde{N}_{bi}^a=\frac{\widehat{%
\partial }\widehat{N}_i^a}{\partial \widehat{y}^b}$ and $\widetilde{N}%
_{bc}^a\left( \widehat{x},\widehat{y}\right) =0.$ For some $\widehat{Y}%
\left( \widehat{u}\right) =\widehat{Y}^i\left( \widehat{u}\right) \widehat{%
\partial }_i+\widehat{Y}^a\left( \widehat{u}\right) \widehat{\partial }_a$
and $\widehat{B}\left( \widehat{u}\right) =\widehat{B}^a\left( \widehat{u}%
\right) \widehat{\partial }_a$ one writes%
$$
\widehat{D}_Y^{(\widetilde{N})}\widehat{B}=\left[ \widehat{Y}^i\left( \frac{%
\widehat{\partial }\widehat{B}^a}{\partial \widehat{x}^i}+\widetilde{N}%
_{bi}^a\widehat{B}^b\right) +\widehat{Y}^b\frac{\widehat{\partial }\widehat{B%
}^a}{\partial \widehat{y}^b}\right] \frac{\widehat{\partial }}{\partial
\widehat{y}^a}.
$$

\item  The d--isoconnection of Berwald type \cite{Berwald}%
$$
\widehat{\Gamma }_{\beta \gamma }^{(B)\alpha }=\left( \widehat{L}_{jk}^i,%
\frac{\widehat{\partial }\widehat{N}_k^a}{\partial \widehat{y}^b},0,\widehat{%
C}_{bc}^a\right) ,
$$
where
$$
\widehat{L}_{.jk}^i\left( \widehat{x},\widehat{y}\right) =\frac 12\widehat{g}%
^{ir}\left( \frac{\widehat{\delta }\widehat{g}_{jk}}{\delta \widehat{x}^k}+%
\frac{\widehat{\delta }\widehat{g}_{kr}}{\delta \widehat{x}^j}-\frac{%
\widehat{\delta }\widehat{g}_{jk}}{\delta \widehat{x}^r}\right) ,\eqno(4.10)
$$
$$
\widehat{C}_{.bc}^a\left( \widehat{x},\widehat{y}\right) =\frac 12\widehat{h}%
^{ad}\left( \frac{\widehat{\partial }\widehat{h}_{bd}}{\partial \widehat{y}^c%
}+\frac{\widehat{\partial }\widehat{h}_{cd}}{\partial \widehat{y}^b}-\frac{%
\partial \widehat{h}_{bc}}{\partial \widehat{y}^d}\right) ,
$$
 is {\bf hv-isometric,} i.e. $\widehat{D}_k^{(B)}\widehat{g}_{ij}=0$
and $\widehat{D}_c^{(B)}\widehat{h}_{ab}=0.$

\item  The isocanonical d--isoconnection $\widehat{{\bf \Gamma }}{\bf ^{(c)}}$
is associated to a isometric $\widehat{{\bf G}}$ of type (3.6) $\widehat{%
\Gamma }_{\beta \gamma }^{(c)\alpha }=(\widehat{L}_{jk}^{(c)i},\widehat{L}%
_{bk}^{(c)a},\widehat{C}_{jc}^{(c)i},\widehat{C}_{bc}^{(c)a}),$ with
coefficients (see (4.10))%
$$
\widehat{L}_{jk}^{(c)i}=\widehat{L}_{.jk}^i,\widehat{C}_{bc}^{(c)a}=\widehat{%
C}_{.bc}^a\eqno(4.11))
$$
$$
\widehat{L}_{bi}^{(c)a}=\widetilde{N}_{bi}^a+\frac 12\widehat{h}^{ac}\left(
\frac{\widehat{\delta }\widehat{h}_{bc}}{\delta \widehat{x}^i}-\widetilde{N}%
_{bi}^d\widehat{h}_{dc}-\widetilde{N}_{ci}^d\widehat{h}_{db}\right) ,
$$
$$
\widehat{C}_{jc}^{(c)i}=\frac 12\widehat{g}^{ik}\frac{\widehat{\partial }%
\widehat{g}_{jk}}{\partial \widehat{y}^c}.
$$
This is a isometric d--isoconnection which satisfies compatibility conditions
$$
\widehat{D}_k^{(c)}\widehat{g}_{ij}=0,\widehat{D}_c^{(c)}\widehat{g}_{ij}=0,%
\widehat{D}_k^{(c)}\widehat{h}_{ab}=0,\widehat{D}_c^{(c)}\widehat{h}_{ab}=0.
$$

\item  We can consider N--adapted {\bf isochristoffel distinguished symbols}
(as in (2.5))%
$$
\widetilde{\Gamma }_{\beta \gamma }^\alpha =\frac 12\widehat{G}^{\alpha \tau
}\left( \widehat{\delta }_\gamma \widehat{G}_{\tau \beta }+\widehat{\delta }%
_\beta \widehat{G}_{\tau \gamma }-\widehat{\delta }_\tau \widehat{G}_{\beta
\gamma }\right) ,\eqno(4.12)
$$
which have the components of d--connection $\widetilde{\Gamma }_{\beta
\gamma }^\alpha =\left( \widehat{L}_{jk}^i,0,0,\widehat{C}_{bc}^a\right) ,$
with $\widehat{L}_{jk}^i$ and $\widehat{C}_{bc}^a$ as in (4.10) if $\widehat{%
G}_{\alpha \beta }$ is taken in the form (4.6).
\end{enumerate}

Arbitrary isolinear isoconnections on a v--isobundle $\widehat{\xi }$ can be
also characterized by theirs deformation isotensors with respect, for
instance, to d--isoconnection (4.12):%
$$
\widehat{\Gamma }_{\beta \gamma }^{(B)\alpha }=\widetilde{\Gamma }_{\beta
\gamma }^\alpha +\widehat{P}_{\beta \gamma }^{(B)\alpha },\widehat{\Gamma }%
_{\beta \gamma }^{(c)\alpha }=\widetilde{\Gamma }_{\beta \gamma }^\alpha +%
\widehat{P}_{\beta \gamma }^{(c)\alpha }
$$
or, in general,%
$$
\widehat{\Gamma }_{\beta \gamma }^\alpha =\widetilde{\Gamma }_{\beta \gamma
}^\alpha +\widehat{P}_{\beta \gamma }^\alpha ,\eqno(4.13)
$$
where $\widehat{P}_{\beta \gamma }^{(B)\alpha },\widehat{P}_{\beta \gamma
}^{(c)\alpha }$ and $\widehat{P}_{\beta \gamma }^\alpha $ are corresponding
deformation d--isotensors of d--iso\-con\-nec\-ti\-ons.

\section{ Isotorsions and Isocurvatures}

The notions of isotorsion and isocurvature were introduced in the ref.
 \cite{Santilli1996} on an isoriemannian spaces.  In this section we
 reformulate these notions on isobundles provided with N--isoconnection and
 d--isoconnection structures.

The {\bf isocurvature} $\widehat{{\bf \Omega }}\,$ of a {\bf nonlinear
 isoconnection} $\widehat{{\bf N}}$ in a v--iso\-bund\-le
 $\widehat{\xi }$ can be
defined as the Nijenhuis tensor isofield $\widehat{N}_v\left( \widehat{X},%
\widehat{Y}\right) $ associated to $\widehat{{\bf N}}{\bf \ }$
(this is an isotopic transform for N--curvature considered, for instance, in
 \cite{MironAnastasiei}):
$$
\widehat{{\bf \Omega }}=\widehat{N}_v={\bf \left[ v\widehat{X},v\widehat{Y}%
\right] +v\left[ \widehat{X},\widehat{Y}\right] -v\left[ v\widehat{X},%
\widehat{Y}\right] -v\left[ \widehat{X},v\widehat{Y}\right] ,}\widehat{{\bf X%
}}{\bf ,}\widehat{{\bf Y}}\in {\cal X}\left( \widehat{\xi }\right)
$$
having this local representation%
$$
\widehat{{\bf \Omega }}=\frac 12\widehat{\Omega }_{ij}^a\widehat{d}%
^i\bigwedge \widehat{d}^j\otimes \widehat{\partial }_a,
$$
where%
$$
\widehat{\Omega }_{ij}^a=\frac{\widehat{\partial }\widehat{N}_i^a}{\partial
\widehat{x}^j}-\frac{\widehat{\partial }\widehat{N}_j^a}{\partial \widehat{x}%
^i}+\widehat{N}_i^b\widetilde{N}_{bj}^a-\widehat{N}_j^b\widetilde{N}_{bi}^a.%
\eqno(5.1)
$$

The {\bf isotorsion} $\widehat{{\bf T}}$ of a {\bf d--isoconnection}
 $\widehat{%
{\bf D}}$ in $\widehat{\xi }$ is defined by the equation%
$$
\widehat{{\bf T}}{\bf \left( \widehat{X},\widehat{Y}\right) =}\widehat{D}_X%
\widehat{{\bf Y}}{\bf -}\widehat{D}_Y\widehat{{\bf X}}{\bf \ -\left[
\widehat{X},\widehat{Y}\right] }. \eqno(5.2)
$$
One holds the following h- and v--decompositions%
$$
\widehat{{\bf T}}{\bf \left( \widehat{X},\widehat{Y}\right) =}\widehat{{\bf T%
}}{\bf \left( h\widehat{X},h\widehat{Y}\right) +}\widehat{{\bf T}}{\bf %
\left( h\widehat{X},v\widehat{Y}\right) +}\widehat{{\bf T}}{\bf \left( v%
\widehat{X},h\widehat{Y}\right) +}\widehat{{\bf T}}{\bf \left( v\widehat{X},v%
\widehat{Y}\right) .}\eqno(5.3)
$$
We consider the projections:
$$
{\bf h}\widehat{{\bf T}}{\bf \left( \widehat{X},\widehat{Y}\right) ,v}%
\widehat{{\bf T}}{\bf \left( h\widehat{X},h\widehat{Y}\right) ,h}\widehat{%
{\bf T}}{\bf \left( h\widehat{X},h\widehat{Y}\right) ,...}
$$
and say that, for instance, ${\bf h}\widehat{{\bf T}}{\bf \left( h\widehat{X}%
,h\widehat{Y}\right) }$ is the h(hh)--isotorsion of $\widehat{{\bf D}},$\\ $%
{\bf v}\widehat{{\bf T}}{\bf \left( h\widehat{X},h\widehat{Y}\right) }$ is
the v(hh)--isotorsion of $\widehat{{\bf D}}$ and so on.

The isotorsion (5.2) is locally determined by five d--tensor isofields,
isotorsions, defined as
$$
\widehat{T}_{jk}^i={\bf h}\widehat{{\bf T}}\left( \widehat{\delta }_k,%
\widehat{\delta }_j\right) \cdot \widehat{d}^i,\quad \widehat{T}_{jk}^a={\bf %
v}\widehat{{\bf T}}\left( \widehat{\delta }_k,\widehat{\delta }_j\right)
\cdot \widehat{\delta }^a,
$$
$$
\widehat{P}_{jb}^i={\bf h}\widehat{{\bf T}}\left( \widehat{\partial }_b,%
\widehat{\delta }_j\right) \cdot \widehat{d}^i,\quad \widehat{P}_{jb}^a={\bf %
v}\widehat{{\bf T}}\left( \widehat{\partial }_b,\widehat{\delta }_j\right)
\cdot \widehat{\delta }^a,
$$
$$
\widehat{S}_{bc}^a={\bf v}\widehat{{\bf T}}\left( \widehat{\partial }_c,%
\widehat{\partial }_b\right) \cdot \widehat{\delta ^a}.
$$
Using formulas (4.2), (4.3), (5.1) and (5.2) we compute in explicit form the
components of isotorsions (5.3) for a d--isoconnection of type (4.8) and (4.9):
$$
\widehat{T}_{.jk}^i=\widehat{T}_{jk}^i=\widehat{L}_{jk}^i-\widehat{L}%
_{kj}^i,\quad \widehat{T}_{ja}^i=\widehat{C}_{.ja}^i,\widehat{T}_{aj}^i=-%
\widehat{C}_{ja}^i,\eqno(5.4)
$$
$$
\widehat{T}_{.ja}^i=0,\widehat{T}_{.bc}^a=\widehat{S}_{.bc}^a=\widehat{C}%
_{bc}^a-\widehat{C}_{cb}^a,
$$
$$
\widehat{T}_{.ij}^a=\frac{\widehat{\delta }N_i^a}{\delta \widehat{x}^j}-%
\frac{\widehat{\delta }\widehat{N}_j^a}{\delta \widehat{x}^i},\quad \widehat{%
T}_{.bi}^a=\widehat{P}_{.bi}^a=\frac{\widehat{\partial }\widehat{N}_i^a}{%
\partial \widehat{y}^b}-\widehat{L}_{.bj}^a,\quad \widehat{T}_{.ib}^a=-%
\widehat{P}_{.bi}^a.
$$

The {\bf isocurvature} $\widehat{{\bf R}}$ of a {\bf d--isoconnection} in $%
\widehat{\xi }$ is defined by the equation
$$
\widehat{{\bf R}}{\bf \left( \widehat{X},\widehat{Y}\right) }\widehat{{\bf Z}%
}{\bf =}\widehat{D}_X\widehat{D}_Y\widehat{{\bf Z}}-\widehat{D}_Y\widehat{D}%
_X\widehat{{\bf Z}}{\bf -}\widehat{D}_{[X,Y]}\widehat{{\bf Z}}{\bf .}%
\eqno(5.5)
$$
One holds the next properties for the h- and v--decompositions of
isocurvature:%
$$
{\bf v}\widehat{{\bf R}}{\bf \left( \widehat{X},\widehat{Y}\right) h}%
\widehat{{\bf Z}}{\bf =0,\ h}\widehat{{\bf R}}{\bf \left( \widehat{X},%
\widehat{Y}\right) v}\widehat{{\bf Z}}{\bf =0,}
$$
$$
\widehat{{\bf R}}{\bf \left( \widehat{X},\widehat{Y}\right) }\widehat{{\bf Z}%
}{\bf =h}\widehat{{\bf R}}{\bf \left( \widehat{X},\widehat{Y}\right) h}%
\widehat{{\bf Z}}{\bf +v}\widehat{{\bf R}}{\bf \left( \widehat{X},\widehat{Y}%
\right) v}\widehat{{\bf Z}}{\bf .}
$$
From (5.5) and the equation $\widehat{{\bf R}}{\bf \left( \widehat{X},%
\widehat{Y}\right) =-}\widehat{{\bf R}}{\bf \left( \widehat{Y},\widehat{X}%
\right) }$ we conclude that the curvature of a d-con\-nec\-ti\-on $\widehat{%
{\bf D}}$ in $\widehat{\xi }$ is completely determined by the following six
d--tensor isofields:%
$$
\widehat{R}_{h.jk}^{.i}=\widehat{d}^i\cdot \widehat{{\bf R}}\left( \widehat{%
\delta }_k,\widehat{\delta }_j\right) \widehat{\delta }_h,~\widehat{R}%
_{b.jk}^{.a}=\widehat{\delta }^a\cdot \widehat{{\bf R}}\left( \widehat{%
\delta }_k,\widehat{\delta }_j\right) \widehat{\partial }_b,\eqno(5.6)
$$
$$
\widehat{P}_{j.kc}^{.i}=\widehat{d}^i\cdot \widehat{{\bf R}}\left( \widehat{%
\partial }_c,\widehat{\partial }_k\right) \widehat{\delta }_j,~\widehat{P}%
_{b.kc}^{.a}=\widehat{\delta }^a\cdot \widehat{{\bf R}}\left( \widehat{%
\partial }_c,\widehat{\partial }_k\right) \widehat{\partial }_b,
$$
$$
\widehat{S}_{j.bc}^{.i}=\widehat{d}^i\cdot \widehat{{\bf R}}\left( \widehat{%
\partial }_c,\widehat{\partial }_b\right) \widehat{\delta }_j,~\widehat{S}%
_{b.cd}^{.a}=\widehat{\delta }^a\cdot \widehat{{\bf R}}\left( \widehat{%
\partial }_d,\widehat{\partial }_c\right) \widehat{\partial }_b.
$$
By a direct computation, using (4.2),(4.3),(4.8),(4.9) and (5.6) we get:
$$
\widehat{R}_{h.jk}^{.i}=\frac{\widehat{\delta }\widehat{L}_{.hj}^i}{\delta
\widehat{x}^h}-\frac{\widehat{\delta }\widehat{L}_{.hk}^i}{\delta \widehat{x}%
^j}+\widehat{L}_{.hj}^m\widehat{L}_{mk}^i-\widehat{L}_{.hk}^m\widehat{L}%
_{mj}^i+\widehat{C}_{.ha}^i\widehat{R}_{.jk}^a,\eqno(5.7)
$$
$$
\widehat{R}_{b.jk}^{.a}=\frac{\widehat{\delta }\widehat{L}_{.bj}^a}{\delta
\widehat{x}^k}-\frac{\widehat{\delta }\widehat{L}_{.bk}^a}{\delta \widehat{x}%
^j}+\widehat{L}_{.bj}^c\widehat{L}_{.ck}^a-\widehat{L}_{.bk}^c\widehat{L}%
_{.cj}^a+\widehat{C}_{.bc}^a\widehat{R}_{.jk}^c,
$$
$$
\widehat{P}_{j.ka}^{.i}=\frac{\widehat{\partial }\widehat{L}_{.jk}^i}{%
\partial \widehat{y}^k}-\left( \frac{\widehat{\partial }\widehat{C}_{.ja}^i}{%
\partial \widehat{x}^k}+\widehat{L}_{.lk}^i\widehat{C}_{.ja}^l-\widehat{L}%
_{.jk}^l\widehat{C}_{.la}^i-\widehat{L}_{.ak}^c\widehat{C}_{.jc}^i\right) +%
\widehat{C}_{.jb}^i\widehat{P}_{.ka}^b,
$$
$$
\widehat{P}_{b.ka}^{.c}=\frac{\widehat{\partial }\widehat{L}_{.bk}^c}{%
\partial \widehat{y}^a}-\left( \frac{\widehat{\partial }\widehat{C}_{.ba}^c}{%
\partial \widehat{x}^k}+\widehat{L}_{.dk}^{c\,}\widehat{C}_{.ba}^d-\widehat{L%
}_{.bk}^d\widehat{C}_{.da}^c-\widehat{L}_{.ak}^d\widehat{C}_{.bd}^c\right) +%
\widehat{C}_{.bd}^c\widehat{P}_{.ka}^d,
$$
$$
\widehat{S}_{j.bc}^{.i}=\frac{\widehat{\partial }\widehat{C}_{.jb}^i}{%
\partial \widehat{y}^c}-\frac{\widehat{\partial }\widehat{C}_{.jc}^i}{%
\partial \widehat{y}^b}+\widehat{C}_{.jb}^h\widehat{C}_{.hc}^i-\widehat{C}%
_{.jc}^h\widehat{C}_{hb}^i,
$$
$$
\widehat{S}_{b.cd}^{.a}=\frac{\widehat{\partial }\widehat{C}_{.bc}^a}{%
\partial \widehat{y}^d}-\frac{\widehat{\partial }\widehat{C}_{.bd}^a}{%
\partial \widehat{y}^c}+\widehat{C}_{.bc}^e\widehat{C}_{.ed}^a-\widehat{C}%
_{.bd}^e\widehat{C}_{.ec}^a.
$$

We note that isotorsions (5.4) and isocurvatures (5.7) can be computed by
particular cases of d--isoconnections when d--isoconnections
(4.11), or (4.12) are used instead of (4.8) and (4.9). The above presented
 formulas are similar
to (2.8),(2.9) and (2.10) being distinguished (in the case of
 locally anisotropic and inhomogeneous  isospaces) by N--isoconnection
structure.

For our further considerations it is useful to compute deformations of
isotorsion (5.2) and isocurvature (5.5) under
 deformations of d--con\-nec\-ti\-ons
(4.13). Putting the splitting (4.13), $\widehat{{\Gamma }}{{^\alpha }_{\beta
\gamma }}={{\tilde \Gamma }_{\cdot \beta \gamma }^\alpha }+\widehat{{P}}{{%
^\alpha }_{\beta \gamma }},$into (5.2) and (5.5) we can express isotorsion $%
\widehat{{T}}{^\alpha }_{\beta \gamma }$ and isocurvature $\widehat{{R}}{{%
_\beta }^\alpha }_{\gamma \delta }$ of a d--isoconnection
 $\widehat{{\Gamma }}{%
^\alpha }_{\beta \gamma }$ as respective deformations of isotorsion
 ${{\tilde T}%
^\alpha }_{\beta \gamma }$ and isotorsion ${\tilde R}_{\beta \cdot \gamma
\delta }^{\cdot \alpha }$ for connection ${{\tilde \Gamma }^\alpha }_{\beta
\gamma }{\quad }:$
$$
{{T^\alpha }_{\beta \gamma }}={{\tilde T}_{\cdot \beta \gamma }^\alpha }+{{%
\ddot T}_{\cdot \beta \gamma }^\alpha }
$$
and
$$
{{{R_\beta }^\alpha }_{\gamma \delta }}={{\tilde R}_{\beta \cdot \gamma
\delta }^{\cdot \alpha }}+{{\ddot R}_{\beta \cdot \gamma \delta }^{\cdot
\alpha }},
$$
where
$$
{{\tilde T}^\alpha }_{\beta \gamma }={{\tilde \Gamma }^\alpha }_{\beta
\gamma }-{{\tilde \Gamma }^\alpha }_{\gamma \beta }+{w^\alpha }_{\gamma
\delta },\qquad {{\ddot T}^\alpha }_{\beta \gamma }={{\ddot \Gamma }^\alpha }%
_{\beta \gamma }-{{\ddot \Gamma }^\alpha }_{\gamma \beta },
$$
and
$$
{{\tilde R}_{\beta \cdot \gamma \delta }^{\cdot \alpha }}={{\delta }_\delta }%
{{\tilde \Gamma }^\alpha }_{\beta \gamma }-{{\delta }_\gamma }{{\tilde
\Gamma }^\alpha }_{\beta \delta }+{{{\tilde \Gamma }^\varphi }_{\beta \gamma
}}{{{\tilde \Gamma }^\alpha }_{\varphi \delta }}-{{{\tilde \Gamma }^\varphi }%
_{\beta \delta }}{{{\tilde \Gamma }^\alpha }_{\varphi \gamma }}+{{\tilde
\Gamma }^\alpha }_{\beta \varphi }{w^\varphi }_{\gamma \delta },
$$
$$
{{\ddot R}_{\beta \cdot \gamma \delta }^{\cdot \alpha }}={{\tilde D}_\delta }%
{{P^\alpha }_{\beta \gamma }}-{{\tilde D}_\gamma }{{P^\alpha }_{\beta \delta
}}+{{P^\varphi }_{\beta \gamma }}{{P^\alpha }_{\varphi \delta }}-{{P^\varphi
}_{\beta \delta }}{{P^\alpha }_{\varphi \gamma }}+{{P^\alpha }_{\beta
\varphi }}{{w^\varphi }_{\gamma \delta }}.
$$

\subsection{Isobianchi and Isoricci Identities}

The isobianchi and isoricci identities were first studied by Santilli
 \cite{Santilli1996} on an isoriemannian space. On spaces with  N--connection
 structures the general formulas for Bianchi and Ricci identities
(for osculator and vector bundles,
 generalized Lagrange and Finsler geometry) have been considered by
 Miron and Anastasiei \cite{MironAnastasiei} and Miron and Atanasiu
 \cite{MironAtanasiu}. We have extended the Miron--Anastasiei--Atanasiu
   constructions for superspaces with local and higher order anisotropy
in refs. \cite{VacaruSupergravity,VacaruNP,VacaruCL}. The purpose of
 this section is to consider distinguished isobianchi and isorichi
 for vector isobundles.

The isotorsion and isocurvature of every linear isoconnection $\widehat{D}$
on a v--isobundle satisfy the following {\bf generalized isobianchi
identities:}
$$
\sum {[(}\widehat{{D}}{_{\widehat{X}}}\widehat{{T}}{)(}\widehat{{Y}}{,}%
\widehat{{Z}}{)-}\widehat{{R}}{(}\widehat{{X}}{,}\widehat{{Y}}{)}\widehat{{Z}%
}{+}\widehat{{T}}{(}\widehat{{T}}{(}\widehat{{X}}{,}\widehat{{Y}}{),}%
\widehat{{Z}}{)]}=0,\eqno(5.8)
$$
$$
\sum {[(}\widehat{{D}}{_{\widehat{X}}}\widehat{{R}}{)(}\widehat{{U}}{,}%
\widehat{{Y}}{,}\widehat{{Z}}{)+}\widehat{{R}}{(}\widehat{{T}}{(}\widehat{{X}%
}{,}\widehat{{Y}}{)}\widehat{{Z}}{)}\widehat{{U}}{]}=0,
$$
where $\sum $ means the respective cyclic sum over $\widehat{X},\widehat{Y},%
\widehat{Z}$ and $\widehat{U}.$ Using the property that
$$
v(\widehat{D}_X\widehat{{R}})(\widehat{U},\widehat{Y},h\widehat{Z})=0,{\quad
}h(\widehat{{D}}{_X}\widehat{R}(\widehat{U},\widehat{Y},v\widehat{Z})=0,
$$
the identities (5.8) become
$$
\sum [{h(}\widehat{{D}}{_X}\widehat{{T}}{)(}\widehat{{Y}}{,}\widehat{{Z}}{)-h%
}\widehat{{R}}{(}\widehat{{X}}{,}\widehat{{Y}}{)}\widehat{{Z}}+\eqno(5.9)
$$
$$
{h}\widehat{{T}}{(h}\widehat{{T}}{(}\widehat{{X}}{,}\widehat{{Y}}{),}%
\widehat{{Z}}{)+h}\widehat{{T}}{(v}\widehat{{T}}{(}\widehat{{X}}{,}\widehat{{%
Y}}{),}\widehat{{Z}}{)]}=0,
$$
$$
\sum {[v(}\widehat{{D}}{_X}\widehat{{T}}{)(}\widehat{{Y}}{,}\widehat{{Z}}{)-v%
}\widehat{{R}}{(}\widehat{{X}}{,}\widehat{{Y}}{)}\widehat{{Z}}{+}
$$
$$
{v}\widehat{{T}}{(h}\widehat{{T}}{(}\widehat{{X}}{,}\widehat{{Y}}{),}%
\widehat{{Z}}{)+v}\widehat{{T}}{(v}\widehat{{T}}{(}\widehat{{X}}{,}\widehat{{%
Y}}{),}\widehat{{Z}}{)]}=0,
$$
$$
\sum {[h(}\widehat{{D}}{_X}\widehat{{R}}{)(}\widehat{{U}}{,}\widehat{{Y}}{,}%
\widehat{{Z}}{)+h}\widehat{{R}}{(h}\widehat{{T}}{(}\widehat{{X}}{,}\widehat{{%
Y}}{),}\widehat{{Z}}{)}\widehat{{U}}{+h}\widehat{{R}}{(v}\widehat{{T}}{(}%
\widehat{{X}}{,}\widehat{{Y}}{),}\widehat{{Z}}{)}\widehat{{U}}{]}=0,
$$
$$
\sum {[v(}\widehat{{D}}{_X}\widehat{{R}}{)(}\widehat{{U}}{,}\widehat{{Y}}{,}%
\widehat{{Z}}{)+v}\widehat{{R}}{(h}\widehat{{T}}{(}\widehat{{X}}{,}\widehat{{%
Y}}{),}\widehat{{Z}}{)}\widehat{{U}}{+v}\widehat{{R}}{(v}\widehat{{T}}{(}%
\widehat{{X}}{,}\widehat{{Y}}{),}\widehat{{Z}}{)}\widehat{{U}}{]}=0.
$$
The local adapted form of these identities is obtained by inserting in (5.9)
the necessary values of triples $(\widehat{X},\widehat{Y},\widehat{Z})$,($=(%
\widehat{{\delta }}{_i},\widehat{{\delta }}{_k},\widehat{{\delta }}{_l}),$
or $(\widehat{{\partial }}{_d},\widehat{{\partial }}{_c},\widehat{{\partial }%
}{_b}),$) and putting successively $\widehat{U}=\widehat{{\delta }}_h$ and $%
\widehat{U}=\widehat{{\partial }}_a.$ Taking into account (4.2),(4.3) and
(5.9) we obtain:
$$
\sum [\widehat{{T}}{_{jk|h}^i+}\widehat{{T}}{^m}_{jk}\widehat{{T}}{^j}_{hm}+%
\widehat{{R}}{^a}_{jk}\widehat{{C}}{^i}_{ha}-\widehat{{R}}{{_j}^i}_{kh}]=0,%
\eqno(5.10)
$$
$$
\sum [\widehat{{R}}{{^a}_{jk{\mid h}}}+\widehat{{T}}{^m}_{jk}\widehat{{R}}{^a%
}_{hm}+\widehat{{R}}{^b}_{jk}\widehat{{P}}{^a}_{hb}]=0,
$$
$$
\widehat{{C}}{^i}_{jb{\mid }k}-\widehat{{C}}{^i}_{kb{\mid }j}-\widehat{{T}}{%
^i}_{jk{\mid }b}+\widehat{{C}}{^m}_{jb}\widehat{{T}}{^i}_{km}-\widehat{C}{^m}%
_{kb}\widehat{{T}}{^i}_{jm}+\widehat{{T}}{^m}_{jk}\widehat{{C}}{^i}_{mb}+
$$
$$
\widehat{{P}}{^d}_{jb}\widehat{{C}}{^i}_{kd}-\widehat{{P}}{^d}_{kb}\widehat{{%
C}}{^i}_{jd}+\widehat{{P}}{{_j}^i}_{kb}-\widehat{{P}}{{_k}^i}_{jb}=0,
$$
$$
\widehat{{P}}{^a}_{jb{\mid }k}-\widehat{{P}}{^a}_{kb{\mid }j}-\widehat{{R}}{%
^a}_{jk\perp b}+\widehat{{C}}{^m}_{jb}\widehat{{R}}{^a}_{km}-\widehat{{C}}{^m%
}_{kb}\widehat{{R}}{^a}_{jm}+
$$
$$
\widehat{{T}}{^m}_{jk}\widehat{{P}}{^a}_{mb}+\widehat{{P}}{^d}_{db}\widehat{{%
P}}{^a}_{kd}-\widehat{{P}}{^d}_{kb}\widehat{{P}}{^a}_{jd}-\widehat{{R}}{{^d}%
_{jk}}\widehat{{S}}{{^a}_{bd}}+\widehat{{R}}{_{b\cdot jk}^{\cdot a}}=0,
$$
$$
\widehat{{C}}{^i}_{jb\perp c}-\widehat{{C}}{^i}_{jc\perp b}+\widehat{{C}}{^m}%
_{jc}\widehat{{C}}{^i}_{mb}-
$$
$$
\widehat{{C}}{^m}_{jb}\widehat{{C}}{^i}_{mc}+\widehat{{S}}{^d}_{bc}\widehat{{%
C}}{^i}_{jd}-\widehat{{S}}{_{j\cdot bc}^{\cdot i}}=0,
$$
$$
\widehat{{P}}{^a}_{jb\perp c}-\widehat{{P}}{^a}_{jc\perp b}+\widehat{{S}}{^a}%
_{bc\mid j}+\widehat{{C}}{^m}_{jc}\widehat{{P}}{^a}_{mb}-\widehat{{C}}{^m}%
_{jb}\widehat{{P}}{^a}_{mc}+
$$
$$
\widehat{{P}}{^d}_{jb}\widehat{{S}}{^a}_{cd}-\widehat{{P}}{^d}_{jc}\widehat{{%
S}}{^a}_{bd}+\widehat{{S}}{^d}_{bc}\widehat{{P}}{^a}_{jd}+\widehat{{P}}{{_b}%
^a}_{jc}-\widehat{{P}}{{_c}^a}_{jb}=0,
$$
$$
\sum [\widehat{{S}}{^a}_{bc\perp d}+\widehat{{S}}{^f}_{bc}\widehat{{S}}{^a}%
_{df}-\widehat{{S}}{{_f}^a}_{cd}]=0,
$$
$$
\sum [\widehat{{R}}{{_k}^i}_{hj\mid l}-\widehat{{T}}{^m}_{hj}\widehat{{R}}{{%
_k}^i}_{lm}-\widehat{{R}}{{^a}_{hj}}\widehat{{P}}{_{k\cdot la}^{\cdot i}}%
]=0,
$$
$$
\sum [\widehat{{R}}{_{d\cdot hj\mid l}^{\cdot a}}-\widehat{{T}}{{^m}_{hj}}%
\widehat{{R}}{_{d\cdot lm}^{\cdot a}}-\widehat{{R}}{{^c}_{hj}}\widehat{{P}}{{%
{_d}^a}_{lc}}]=0,
$$
$$
\widehat{{P}}{_{k\cdot jd\mid l}^{\cdot i}}-\widehat{{P}}{_{k\cdot ld\mid
j}^{\cdot i}}+{{R_k}^i}_{lj\perp d}+{C^m}_{ld}{{R_k}^i}_{jm}-{C^m}_{jd}{{R_k}%
^i}_{lm}-
$$
$$
\widehat{{T}}{^m}_{jl}\widehat{{P}}{_{k\cdot md}^{\cdot i}}+\widehat{{P}}{^a}%
_{ld}\widehat{{P}}{_{k\cdot jl}^{\cdot i}}-\widehat{{P}}{{^a}_{jd}}\widehat{{%
P}}{_{k\cdot la}^{\cdot i}}-\widehat{{R}}{{^a}_{jl}}\widehat{{S}}{_{k\cdot
ad}^{\cdot i}}=0,
$$
$$
\widehat{{P}}{{_c}^a}_{jd\mid l}-\widehat{{P}}{{_c}^a}_{ld\mid j}+\widehat{{R%
}}{_{c\cdot lj\mid d}^{\cdot a}}+\widehat{{C}}{^m}_{ld}\widehat{{R}}{{_c}^a}%
_{jm}-\widehat{{C}}{^m}_{jd}\widehat{{R}}{{_c}^a}_{lm}-
$$
$$
\widehat{{T}}{^m}_{jl}\widehat{{P}}{{_c}^a}_{md}+\widehat{{P}}{^f}_{ld}%
\widehat{{P}}{{_c}^a}_{jf}-\widehat{{P}}{^f}_{jd}\widehat{{P}}{{_c}^a}_{lf}-%
\widehat{{R}}{^f}_{jl}\widehat{{S}}{{_c}^a}_{fd}=0,
$$
$$
\widehat{{P}}{_{k\cdot jd\perp c}^{\cdot i}}-\widehat{{P}}{_{k\cdot jc\perp
d}^{\cdot i}}+\widehat{{S}}{{_k}^i}_{dc\mid j}+\widehat{{C}}{^m}_{jd}%
\widehat{{P}}{_{k\cdot mc}^{\cdot i}}-\widehat{{C}}{^m}_{jc}\widehat{{P}}{%
_{k\cdot md}^{\cdot i}}+
$$
$$
\widehat{{P}}{^a}_{jc}\widehat{{S}}{_{k\cdot da}^{\cdot i}}-\widehat{{P}}{^a}%
_{jd}\widehat{{S}}{_{k\cdot ca}^{\cdot i}}+\widehat{{S}}{^a}_{cd}\widehat{{P}%
}{_{k\cdot ja}^{\cdot i}}=0,
$$
$$
\widehat{{P}}{{_b}^a}_{jd\perp c}-\widehat{{P}}{{_b}^a}_{jc\perp d}+\widehat{%
{S}}{{_b}^a}_{cd\mid j}+\widehat{{C}}{^m}_{jd}\widehat{{P}}{{_b}^a}_{mc}-
$$
$$
\widehat{{C}}{^m}_{jc}\widehat{{P}}{{_b}^a}_{md}+\widehat{{P}}{^f}_{jc}%
\widehat{{S}}{{_b}^a}_{df}-\widehat{{P}}{^f}_{jd}\widehat{{S}}{{_b}^a}_{cf}+%
\widehat{{S}}{^f}_{cd}\widehat{{P}}{{_b}^a}_{jf}=0,
$$
$$
\sum_{[b,c,d]}\widehat{{S}}{{{_k}^i}_{bc\perp d}-}\widehat{{S}}{{^a}_{bc}}%
\widehat{{S}}{{{{_k}^i}_{da}}}
$$
$$
\sum_{[b,c,d]}{[}\widehat{{S}}{{{_f}^a}_{bc\perp d}-}\widehat{{S}}{{^e}_{bc}}%
\widehat{{S}}{{{{_f}^a}_{de}}]}=0,
$$
where, for instance, ${\sum_{[b,c,d]}}$ means the cyclic sum over indices $%
b,c,d.$

Identities (5.10) are isotopic generalizations of the corresponding formulas
presented in \cite{MironAnastasiei}, or equivalently, an extension of
 Santilli's  \cite{Santilli1996} formulas to the case of d--isoconnections.

As a consequence of a corresponding rearrangement of (5.9) we obtain the
{\bf isoricci identities} (for simplicity we establish them only for
distinguished vector isofields, although they may be written for every
distinguished tensor isofield):
$$
\widehat{{D}}{_{[X}^{(h)}}\widehat{{D}}{_{Y\}}^{(h)}}h\widehat{Z}=\widehat{R}%
(h\widehat{X},h\widehat{Y})h\widehat{Z}+\widehat{{D}}{_{[hX,hY\}}^{(h)}}h%
\widehat{Z}+\widehat{{D}}{_{[hX,hY\}}^{(v)}}h\widehat{Z},\eqno(5.11)
$$
$$
\widehat{{D}}{_{[X}^{(v)}}{D_{Y\}}^{(h)}}h\widehat{Z}=\widehat{R}(v\widehat{X%
},h\widehat{Y})h\widehat{Z}+\widehat{{D}}{_{[vX,hY\}}^{(h)}}h\widehat{Z}+%
\widehat{{D}}{_{[vX,hY\}}^{(v)}}h\widehat{Z},
$$
$$
\widehat{{D}}{_{[X}^{(v)}}\widehat{{D}}{_{Y\}}^{(v)}}h\widehat{Z}=\widehat{R}%
(v\widehat{X},v\widehat{Y})h\widehat{Z}+\widehat{{D}}{_{[vX,vY\}}^{(v)}}h%
\widehat{Z}
$$
and
$$
\widehat{{D}}{_{[X}^{(h)}}\widehat{{D}}{_{Y\}}^{(h)}}v\widehat{Z}=\widehat{R}%
(h\widehat{X},h\widehat{Y})v\widehat{Z}+{D_{[hX,hY\}}^{(h)}}v\widehat{Z}+{%
D_{[hX,hY\}}^{(v)}}v\widehat{Z},\eqno(5.12)
$$
$$
\widehat{{D}}{_{[X}^{(v)}}\widehat{{D}}{_{Y\}}^{(h)}}v\widehat{Z}=\widehat{R}%
(v\widehat{X},h\widehat{Y})v\widehat{Z}+\widehat{{D}}{_{[vX,hY\}}^{(v)}}v%
\widehat{Z}+\widehat{{D}}{_{[vX,hY\}}^{(v)}}v\widehat{Z},
$$
$$
\widehat{{D}}{_{[X}^{(v)}}\widehat{{D}}{_{Y\}}^{(v)}}v\widehat{Z}=\widehat{R}%
(v\widehat{X},v\widehat{Y})v\widehat{Z}+\widehat{{D}}{_{[vX,vY\}}^{(v)}}v%
\widehat{Z}.
$$
For $\widehat{X}=\widehat{{X}}{^i}(\widehat{u}){\frac{\widehat{\delta }}{%
\delta \widehat{x}^i}}+\widehat{{X}}{^a}(\widehat{u})\frac{\widehat{\partial
}}{\partial \widehat{x}^a}$ and (4.2),(4.3),(5.4) and (5.7) we can express
respectively identities (5.11) and (5.12) in this form:
$$
\widehat{{X}}{^a}_{\mid k\mid l}-\widehat{{X}}{^a}_{\mid l\mid k}=\widehat{{R%
}}{{{_B}^a}_{kl}}\widehat{{X}}{^b}-\widehat{{T}}{^h}_{kl}\widehat{{X}}{^a}%
_{\mid h}-\widehat{{R}}{^b}_{kl}\widehat{{X}}{^a}_{\perp b},
$$
$$
\widehat{{X}}{^i}_{\mid k\perp d}-\widehat{{X}}{^i}_{\perp d\mid k}=\widehat{%
{P}}{_{h\cdot kd}^{\cdot i}}\widehat{{X}}{^h}-\widehat{{C}}{^h}_{kd}\widehat{%
{X}}{^i}_{\mid h}-\widehat{{P}}{^a}_{kd}\widehat{{X}}{^i}_{\perp a},
$$
$$
\widehat{{X}}{^i}_{\perp b\perp c}-\widehat{{X}}{^i}_{\perp c\perp b}=%
\widehat{{S}}{_{h\cdot bc}^{\cdot i}}\widehat{{X}}{^h}-\widehat{{S}}{^a}_{bc}%
\widehat{{X}}{^i}_{\perp a}
$$
and
$$
\widehat{{X}}{^a}_{\mid k\mid l}-\widehat{{X}}{^a}_{\mid l\mid k}=\widehat{{R%
}}{{_b}^a}_{kl}\widehat{{X}}{^b}-\widehat{{T}}{^h}_{kl}\widehat{{X}}{^a}%
_{\mid h}-\widehat{{R}}{^b}_{kl}\widehat{{X}}{^a}_{\perp b},
$$
$$
\widehat{{X}}{^a}_{\mid k\perp b}-\widehat{{X}}{^a}_{\perp b\mid k}=\widehat{%
{P}}{{_b}^a}_{kc}\widehat{{X}}{^c}-\widehat{{C}}{^h}_{kb}\widehat{{X}}{^a}%
_{\mid h}-\widehat{{P}}{^d}_{kb}\widehat{{X}}{^a}_{\perp d},
$$
$$
\widehat{{X}}{^a}_{\perp b\perp c}-\widehat{{X}}{^a}_{\perp c\perp b}=%
\widehat{{S}}{{_d}^a}_{bc}\widehat{{X}}{^d}-\widehat{{S}}{^d}_{bc}\widehat{{X%
}}{^a}_{\perp d}.
$$

For some considerations it is useful to use an alternative way of definition
isotorsion (5.2) and isocurvature (5.5) by using the commutator
$$
\widehat{\Delta }_{\alpha \beta }\doteq \widehat{\nabla }_\alpha \widehat{%
\nabla }_\beta -\widehat{\nabla }_\beta \widehat{\nabla }_\alpha =2\widehat{%
\nabla }_{[\alpha }\widehat{\nabla }_{\beta ]}.\eqno(5.13)
$$
For components (5.13) of d--isotorsion we have
$$
\widehat{\Delta }_{\alpha \beta }\widehat{f}=\widehat{T}_{.\alpha \beta
}^\gamma \widehat{\nabla }_\gamma \widehat{f}
$$
for every scalar function $\widehat{f}\,$ on $\widehat{\xi }.$ Curvature can
be introduced as an operator acting on arbitrary d--isovector $\widehat{V}%
^\delta :$
$$
(\widehat{\Delta }_{\alpha \beta }-\widehat{T}_{.\alpha \beta }^\gamma
\widehat{\nabla }_\gamma )\widehat{V}^\delta =\widehat{R}_{~\gamma .\alpha
\beta }^{.\delta }\widehat{V}^\gamma
$$
(in this work we are following conventions similar to Miron and Anastasiei
\cite{MironAnastasiei} on d--isotensors; we can obtain corresponding Penrose
and Rindler abstract index formulas \cite{Penrose} just for a trivial
N--connection structure and by changing denotations for components of
isotorsion and isocurvature in this manner:\ $T_{.\alpha \beta }^\gamma
\rightarrow T_{\alpha \beta }^{\quad \gamma }$ and $R_{~\gamma .\alpha \beta
}^{.\delta }\rightarrow R_{\alpha \beta \gamma }^{\qquad \delta }).$

\subsection{Structure Equations of a  d--Isoconnection}

Let us, for instance, consider d--tensor isofield:
$$
\widehat{t}=\widehat{{t}}{_a^i}\widehat{{\delta }}_I{\otimes }\widehat{{%
\delta }}{^a}.
$$
We introduce the so--called d--connection 1--forms ${\omega }_j^i$ and ${{%
\tilde \omega }_b^a}$ as%
$$
\widehat{D}\widehat{t}=(\widehat{D}\widehat{{t}}{_a^i})\widehat{{\delta }}_I{%
\otimes }\widehat{{\delta }}^a
$$
with
$$
\widehat{D}\widehat{t}_a^i=\widehat{d}\widehat{t}_a^i+{\omega }_j^i\widehat{{%
t}}{_a^j}-{{\tilde \omega }_a^b}\widehat{{t}}{_b^i}=\widehat{t}_{a\mid j}^i%
\widehat{{d}}\widehat{{x}}{^j}+\widehat{t}_{a\perp b}^I\widehat{{\delta }}%
\widehat{y}^b.
$$
For the d--isoconnection 1--forms of a
 d--isoconnection $\widehat{D}$ on $\widehat{%
\xi }$ defined by ${{\omega }_j^i}$ and ${{\tilde \omega }_b^a}$ one holds
the following {\bf structure isoequations: }%
$$
d(\widehat{{d}}^i)-\widehat{{d}}{^h}\wedge {\omega }_h^i=-\widehat{{\Omega }}%
,
$$
$$
d{(}\widehat{{\delta }}{{^a})}-\widehat{{\delta }}{^a}\wedge {\omega _b^a}=-%
\widehat{{\Omega }}{^a},
$$
$$
d{{\omega }_j^i}-{{\omega }_j^h}\wedge {{\omega }_h^i}=-\widehat{{\Omega }}{%
_j^i},
$$
$$
d{\omega _b^a}-{\omega _b^c}\wedge {\omega _c^a}=-\widehat{{\Omega }}{_b^a},
$$
in which the isotorsion 2--forms $\widehat{{\Omega }}^i$ and $\widehat{{%
\Omega }}{^i}$ are given respectively by formulas:
$$
\widehat{{\Omega }}{^i}={\frac 12}\widehat{{T}}{^i}_{jk}\widehat{{d}}{^j}%
\wedge \widehat{{d}}{^k}+{\frac 12}\widehat{{C}}{^i}_{jk}\widehat{{d}}{^j}%
\wedge \widehat{{\delta }}{^c},
$$
$$
\widehat{{\Omega }}{^a}={\frac 12}\widehat{{R}}{^a}_{jk}\widehat{{d^j}}%
\wedge \widehat{{d}}{^k}+{\frac 12}\widehat{{P}}{^a}_{jc}\widehat{{d}}{^j}%
\wedge \widehat{{\delta }}{^c}+{\frac 12}\widehat{{S}}{^s}_{bc}\widehat{{%
\delta }}{^b}\wedge \widehat{{\delta }}{^c},
$$
and
$$
\widehat{{\Omega }}{_j^i}={\frac 12}\widehat{{R}}{{_j}^i}_{kh}\widehat{{d}}{%
^k}\wedge \widehat{{d}}{^h}+{\frac 12}\widehat{P}{_{j\cdot kc}^{\cdot i}}%
\widehat{{d}}{^k}\wedge \widehat{{\delta }}{^c}+{\frac 12}\widehat{S}{%
_{j\cdot kc}^{\cdot i}}\widehat{{\delta }}{^b}\wedge \widehat{{\delta }}{^c}%
,
$$
$$
\widehat{\Omega }{_b^a}={\frac 12}\widehat{{R}}{_{b\cdot kh}^{\cdot a}}%
\widehat{{d}}{^k}\wedge \widehat{{d}}{^h}+\frac 12\widehat{P}_{b.kc}^{.a}%
\widehat{d}^k\wedge \widehat{\delta }^c+\frac 12\widehat{S}_{b.cd}^{.a}%
\widehat{{\delta }}{^c}\wedge \widehat{{\delta }}{^d}
$$
The just presented formulas are very similar to those for usual locally
anisotropic spaces \cite{MironAnastasiei} but in our case they are written
for isotopic values and generalize the isoriemannian
Santilli's formulas \cite{Santilli1996}.

\section{The Isogeometry of Tangent Iso\-bundles}

The aim of this section is to formulate some results in the isogeometry of
tangent isobundle, t--isobundle, $\widehat{TM}$ and to use them in
order to develop the geometry of Finsler and Lagrange isospaces.

All results presented in the preceding section  on v--isobundles
provided with N--isoconnection, d--isoconnection and isometric
 structures hold
good for $\widehat{TM}.$ In this case the dimension of the base isospace and
of typical isofibre coincides and we can write locally, for instance,
isovectors as
$$
\widehat{X}=\widehat{{X}}{^i}\widehat{{\delta }}_i+\widehat{{Y}}{^i}\widehat{%
{\partial }}_i=\widehat{X}^i\widehat{{\delta }}_i+\widehat{Y}^{(i)}\widehat{{%
\partial }}_{(i)},
$$
where $\widehat{u}^\alpha =(\widehat{x}^i,\widehat{y}^j)=(\widehat{x}^i,%
\widehat{y}^{(j)}).$

On t-isobundles we can define a global map
$$
\widehat{J}:{\cal X}\left( \widehat{TM}\right) \to {\cal X}\left( \widehat{TM%
}\right) \eqno(6.1)
$$
which does not depend on N--isoconnection structure:
$$
\widehat{J}({\frac{\widehat{\delta }}{\delta \widehat{x}^i}})={\frac{%
\widehat{\partial }}{\partial \widehat{y}^i}}
$$
and%
$$
\widehat{J}({\frac{\widehat{\partial }}{\partial \widehat{y}^i}})=0.
$$
This endomorphism is called the {\bf natural (or canonical) almost tangent
isostructure} on $\widehat{TM}$; it has the properties:
$$
1)\widehat{J}^2=0,{\quad }2)Im\widehat{J}=Ker\widehat{J}=V\widehat{TM}
$$
and 3) the {\bf Nigenhuis isotensor,}
$$
{N_J}(\widehat{X},\widehat{Y})=[J\widehat{X},J\widehat{Y}\}-J[J\widehat{X},%
\widehat{Y}\}-J[\widehat{X},J\widehat{Y}]
$$
$$
(\widehat{X},\widehat{Y}\in {\cal X}\left( \widehat{TM}\right) )
$$
identically vanishes, i.e. the natural almost tangent isostructure $J$ on $%
\widehat{TM}$ is isointegrable.

\subsection{Notions of Generalized Isolagrange, Isolagrange and
 Isofinsler Spaces}

Let $\widehat{M}$ be a isosmooth $(2n)$--dimensional isomanifold and $(%
\widehat{TM},{\tau },\widehat{M})$ its t--isobundle. For isospaces we
 define
a {\bf generalized isolagrange space,} GIL--space,  as a pair ${G}%
\widehat{{L}}^{n,m}=(\widehat{M},\widehat{g}_{ij}(\widehat{x},\widehat{y}))$%
, where $\widehat{g}_{ij}(\widehat{x},\widehat{y})$ is a d--tensor isofield
on ${\tilde TM}=\widehat{TM}-\{0\},$ of isorank $(2n),$ and is called as the
fundamental d--isotensor, or metric d--isotensor, of GIL--space.

Let denote as a normal d--isoconnection that defined by using $N$ and being
adapted to the almost tangent isostructure (6.1) as $\widehat{D}{\Gamma }=(%
\widehat{{L}}{^a}_{jk},\widehat{{C}}{^a}_{jk}).$ This d--isoconnection is
compatible with isometric $\widehat{g}_{ij}(\widehat{x},\widehat{y})$ if $%
\widehat{g}_{ij\mid k}=0$ and $\widehat{g}_{ij\perp k}=0.$

There exists an unique d--isoconnection $C\widehat{\Gamma }(N)$ which is
compatible with $\widehat{g}_{ij}{(}\widehat{{u}}{)}$ and has vanishing
isotorsions $\widehat{{T}}{^i}_{jk}$ and $\widehat{{S}}{^i}_{jk}$ (see
formulas (5.4) rewritten for t--isobundles). This isoconnection, depending only
on $\widehat{g}_{ij}{(}\widehat{{u}}{)}$ and $\widehat{{N}}_j^i{(}\widehat{{u%
}}{)}$ is called the canonical metric d--isoconnection of GIL--space. It has
coefficients
$$
\widehat{{L}}{^i}_{jk}={\frac 12}\widehat{{g}}{^{ih}}(\widehat{{\delta }}_j%
\widehat{{g}}{_{hk}}+\widehat{{\delta }}_h\widehat{{g}}{_{jk}}-\widehat{{%
\delta }}_h\widehat{{g}}{_{jk}}),\eqno(6.2)
$$
$$
\widehat{{C}}{^i}_{jk}={\frac 12}\widehat{{g}}{^{ih}}(\widehat{{\partial }}_j%
\widehat{{g}}{_{hk}}+\widehat{{\partial }}_h\widehat{{g}}{_{jk}}-\widehat{{%
\partial }}_h\widehat{{g}}{_{jk}}).
$$
Of course, metric d--isoconnections different from $C\widehat{\Gamma }(N)$ may
be found. For instance, there is a unique normal d--isoconnection $\widehat{D}%
\Gamma (N)=({\bar L}_{\cdot jk}^i,{\bar C}_{\cdot jk}^i)$ which is metric
and has a priori given isotorsions $\widehat{{T}}{^i}_{jk}$ and $\widehat{{S}%
}{^i}_{jk}.$ The coefficients of $\widehat{D}\Gamma (N)$ are the following
ones:
$$
{\bar L}_{\cdot jk}^i=\widehat{{L}}{^i}_{jk}-\frac 12\widehat{g}^{ih}(%
\widehat{g}_{jr}\widehat{{T}}{^r}_{hk}+\widehat{g}_{kr}\widehat{{T}}{^r}%
_{hj}-\widehat{g}_{hr}\widehat{{T}}{^r}_{kj}),
$$
$$
{\bar C}_{\cdot jk}^i=\widehat{{C}}{^i}_{jk}-\frac 12\widehat{g}^{ih}(%
\widehat{g}_{jr}\widehat{{S}}{^r}_{hk}+\widehat{g}_{kr}\widehat{{S}}{^r}%
_{hj}-\widehat{g}_{hr}\widehat{{S}}{^r}_{kj}),
$$
where $\widehat{{L}}{^i}_{jk}$ and $\widehat{{C}}{^i}_{jk}$ are the same as
for the $C\widehat{\Gamma }(N)$--isoconnection (6.2).

The Lagrange spaces were introduced in order to geometrize the concept of
Lagrangian in mechanics (the Lagrange geometry is studied in details, see
also basic references, in Miron and Anastasiei
\cite{MironAnastasiei}). For isospaces we present
this generalization:

A {\bf isolagrange space}, IL--space, $\widehat{L}^n=(\widehat{M},\widehat{%
g}_{ij}),$ is defined as a particular case of GIL--space when the d--isometric
on $\widehat{M}$ can be expressed as
$$
\widehat{g}_{ij}{(}\widehat{{u}}{)}={\frac 12}{\frac{\widehat{{\partial }}^2%
{\cal L}}{{\partial }\widehat{{y}}{{^i}{\partial }}\widehat{{y}}{^j}}},%
\eqno(6.3)
$$
where ${\cal L}:\widehat{TM}\to \widehat{\Lambda },$ is a isodifferentiable
function called a iso--Lagrangian on $\widehat{M}$.

Now we consider the isotopic extension of the Finsler space:

A {\bf isofinsler isometric} on $\widehat{M}$ is a function $F_S:\widehat{TM}%
\to \widehat{\Lambda }$ having the properties:

1. The restriction of $F_S$ to ${\tilde {TM}}=\widehat{TM}\setminus \{0\}$
is of the class $G^\infty $ and F is only isosmooth on the image of the null
cross--section in the t--isobundle to $\widehat{M}$.

2. The restriction of $\widehat{F}$ to ${\tilde {TM}}$ is positively
homogeneous of degree 1 with respect to ${(}\widehat{{y}}{^i)}$, i.e. $%
\widehat{F}(\widehat{x},\widehat{{\lambda }}\widehat{y})=\widehat{{\lambda }}%
\widehat{F}(\widehat{x},\widehat{y}),$ where $\widehat{{\lambda }}$ is a
real positive number.

3. The restriction of $\widehat{F}$ to the even subspace of $\tilde {TM}$ is
a positive function.

4. The quadratic form on ${\Lambda }^n$ with the coefficients
$$
\widehat{g}_{ij}{(}\widehat{{u}}{)}={\frac 12}{\frac{\widehat{{\partial }}^2%
\widehat{F}^2}{{\partial }\widehat{{y}}{{^i}{\partial }}\widehat{{y}}{^j}}}%
\eqno(6.4)
$$
defined on $\tilde {TM}$ is nondegenerate.

A pair $\widehat{F}^n=(\widehat{M},\widehat{F})$ which consists from a
continuous isomanifold $\widehat{M}$ and a isofinsler isometric is called a
{\bf isofinsler space}, IF--space.

It's obvious that IF--spaces form a particular class of IL--spaces with
iso-Lagrangian ${\cal L}=\widehat{{F}}{^2}$ and a particular class of
GIL--spaces with metrics of type (6.4).

For a IF--space we can introduce the isotopic variant of nonlinear Cartan
connection \cite{MironAnastasiei}:
$$
\widehat{N}_j^i{(}\widehat{{x}}{,}\widehat{{y}}{)}={\frac{\widehat{\partial }%
}{\partial \widehat{y}^j}}\widehat{G}^{*I},
$$
where
$$
\widehat{G}^{*i}={\frac 14}\widehat{g}^{*ij}({\frac{\widehat{{\partial }}^2{%
\varepsilon }}{{\partial }\widehat{{y}}{^i}{\partial }\widehat{{x}}{^k}}}%
\widehat{{y}}{^k}-{\frac{\widehat{\partial }{\varepsilon }}{\partial
\widehat{x}^j}}),{\quad }{\varepsilon }{(}\widehat{{u}}{)}=\widehat{g}_{ij}{(%
}\widehat{{u}}{)}\widehat{y}^i\widehat{y}^j,
$$
and $\widehat{g}^{*ij}$ is inverse to $\widehat{g}_{ij}^{*}{(}\widehat{{u}}{)%
}={\frac 12}{\frac{\widehat{{\partial }}^2\varepsilon }{{\partial }\widehat{{%
y}}{{^i}{\partial }}\widehat{{y}}{^j}}}.$ In this case the coefficients of
canonical metric d--isoconnection (6.2) gives the isotopic variants of
coefficients of the Cartan connection of Finsler spaces. A similar remark
applies to the isolagrange spaces.

\subsection{The Isotopic Almost Hermitian Model of the GIL--Space}

Consider a GIL--space endowed with the canonical metric d--isoconnection $C%
\widehat{\Gamma }(N).$ Let $\widehat{{\delta }}_\alpha =(\widehat{{\delta }}%
_\alpha ,\widehat{{\dot \partial }}_I)$ be a usual adapted frame (4.2) on TM
and $\widehat{{\delta }}^\alpha =(\widehat{{\partial }}^I,\widehat{{\dot
\delta }}^I)$ its dual, see (4.3). The linear operator
$$
\widehat{F}:\Xi ({\tilde {TM}})\to \Xi ({\tilde {TM}}),
$$
acting on $\widehat{{\delta }}_\alpha $ by $\widehat{F}(\widehat{{\delta }}%
_i)=-\widehat{{\partial }}_i,\widehat{F}(\widehat{{\dot \partial }}_i)=%
\widehat{{\delta }}_i,$ defines an almost complex isostructure on ${T}\widehat{{%
M}}.$ We shall obtain a complex isostructure if and only if the even
component of the horizontal distribution $\widehat{N}$ is integrable. For
isospaces, in general with even and odd components, we write the isotopic
almost Hermitian property (almost Hermitian isostructure) as
$$
\widehat{{F}}{_\beta ^\alpha }\widehat{{F}}{_\delta ^\beta }=-{{\delta }%
_\beta ^\alpha }.
$$

The isometric $\widehat{g}_{ij}{(}\widehat{{x}}{,}\widehat{{y}}{)}$ on
GIL--spaces induces on $\dot {T\widehat{M}}$ the following isometric:
$$
\widehat{G}=\widehat{g}_{ij}{(}\widehat{{u}}{)}\widehat{d}\widehat{x}%
^i\otimes \widehat{d}\widehat{x}^j+\widehat{g}_{ij}{(}\widehat{{u}}{)}%
\widehat{{\delta }}\widehat{{y}}{^i}\otimes \widehat{{\delta }}\widehat{{y}}{%
^j}.\eqno(6.5)
$$
We can verify that pair $(\widehat{G},\widehat{F})$ is an almost Hermitian
isostructure on ${\dot {T\widehat{M}}}$ with the associated supersymmetric
2--form
$$
\widehat{\theta }=\widehat{g}_{ij}{(}\widehat{{x}}{,}\widehat{{y}}{)}%
\widehat{{\delta }}\widehat{{y}}{^i}\wedge \widehat{d}\widehat{x}^j.
$$

The almost Hermitian isospace $\widehat{H}^{2n}=(T\widehat{M},\widehat{G},%
\widehat{F}),$ provided with a isometric of type (6.5) is called the lift on $T%
\widehat{M}$, or the almost Hermitian isomodel, of GIL--space $G\widehat{L}%
^n.$ We say that a linear isoconnection $\widehat{D}$ on ${\dot {T\widehat{M}}}$
is almost Hermitian isotopic of Lagrange type if it preserves by parallelism
the vertical distribution $V$ and is compatible with the almost Hermitian
isostructure $(\widehat{G},\widehat{F})$, i.e.
$$
\widehat{D}_X\widehat{G}=0,\quad \widehat{D}_X\widehat{F}=0,\eqno(6.6)
$$
for every $X\in \widehat{{\cal X}}\left( T\widehat{M}\right) .$

There exists an unique almost Hermitian isoconnection of Lagrange type $%
\widehat{D}^{(c)}$ having h(hh)- and v(vv)--isotorsions equal to zero. We
can prove (similarly as in \cite{MironAnastasiei}) that coefficients ${(}%
\widehat{{L}}{{^i}_{jk},}\widehat{{C}}{{^i}_{jk})}$ of $\widehat{D}^{(c)}$
in the adapted basis $(\widehat{{\delta }}_i,\widehat{{\dot \delta }}_j)$
are just the coefficients (6.2) of the canonical metric d--isoconnection $C%
\widehat{\Gamma }(N)$ of the GIL--space $G\widehat{L}^n.$ Inversely, we can
say that $C\widehat{\Gamma }(N)$--connection determines on ${\tilde {TM}}$
and isotopic almost Hermitian connection of Lagrange type with vanishing
h(hh)- and v(vv)-isotorsions. If instead of GIL--space isometric $\widehat{g}%
_{ij}$ in (6.4) the isolagrange (or isofinsler) isometric (6.2) (or (6.3)) is
taken, we obtain the almost Hermitian isomodel of isolagrange (or isofinsler)
isospaces $\widehat{L}^n$ (or $\widehat{{F}}{^n}).$

We note that the natural compatibility conditions (6.6) for the isometric
(6.5) and $C\widehat{\Gamma }(N)$--con\-nec\-ti\-ons on $\widehat{H}^{2n}$%
--spaces plays an important role for developing physical models on
la--isospaces. In the case of usual locally anisotropic spaces geometric
constructions and d--covariant calculus are very similar to those for the
Riemann and Einstein--Cartan spaces. This is exploited for formulation in a
selfconsistent manner the theory of spinors on la--spaces \cite{VacaruJMP},
for introducing a geometric background for locally anisotropic Yang--Mills
and gauge like gravitational interactions \cite{VacaruGoncharenko} and for
extending the theory of stochastic processes and diffusion to the case of
locally anisotropic spaces and interactions on such spaces \cite
{VacaruStochastics}. In a similar manner we can introduce N--lifts to v- and
t--isobundles in order to investigate isotopic gravitational la--models.

\section{Isogravity on Locally Anisotropic and Inhomogeneous Isospa\-ces}

The conventional Riemannian geometry can be generally assumed to be exactly
valid for the exterior gravitational problem in vacuum where  bodies can
 be well approximated as being massive points, thus implying the validity
 of conventional and calculus.

On the contrary, there have been serious doubt dating back to E. Cartan
 on the same exact validity of the Riemannian geometry for interior
gravitational problem because the latter imply internal effects which are
 arbitrary nonlinear in the velocities and other variables, nonlocal
 integral and of general non--(first)--order Lagrangian type.

Santilli \cite{Santilli1978,Santilli1979,Santilli1980,Santilli1996}
 constructed his isoriemannian geometry and proposed the related
 isogravitation theory precisely to resolve the latter shortcoming. In
 fact, the  isometric acquires an arbitrary functional; dependence thus
 being   able to   represent directly the locally anisotropic and
 inhomogeneous character of interior gravitational problems.

A remarkable aspect of the latter advances is that they were achieved by
 preserving the  abstract geometric axioms of the exterior gravitation.
 In fact, exterior and interior gravitation are unified in the above
geometric approach and are merely differentiated by the selected unit,
 the trivial value $I=diag(1,1,1,1)$ yielding the conventional gravitation
 in vacuum while more general realization of the unit yield interior
 conditions under the same abstract axioms (see ref.
\cite{Kadeisvili} for an independent study).

A number of applications of the isogeometries for interior problems have
already been  identified, such as (see ref. \cite{Santilli1980} for an
 outline): the representation of the local variation of the speed of light
 within physical media such as atmospheres or chromospheres; the
 representation of the large difference between cosmological redshift
 between certain quasars and their associated galaxies when physically
 connected according to spectroscopic evidence; the initiation of the
 study of the origin  of the   gravitation via its identification with
 the field originating the mass of elementary  constituents.

As we have shown \cite{VacaruAP,VacaruNP} the low energy limits of string
 and superstring theories  give  also rise
 to models of (super)field interactions
 with locally anisotropic and even higher order  anisotropic interactions.
The N--connection field can be treated as a corresponding nonlinear
 gauge field managing the dynamics of ''step by step'' splitting (reduction)
 of higher dimensional spaces to lower dimensional ones. Such (super)string
 induced (super)gravitational models have a generic local anisotropy
 and, in consequence, a more sophisticate  form of field equations and
 conservation laws and of corresponding theirs stochastic and quantum
 modifications. Perhaps similar considerations are in right for isotopic
 versions of sting theories. That it is why we are interested in a study
 of models of isogravity with nonvanishing nonlinear isoconnection,
 distinguished isotorsion and, in general, non--isometric fields.

To begin our presentation let us consider
 a v--isobundle $\widehat{\xi }=(\widehat{E},{\pi },\widehat{M})$
provided with some compatible nonlinear isoconnection $\widehat{N}$,
d--isoconnection $\widehat{D}$ and isometric $\widehat{G}$ structures.For a
locally N--adapted isoframe we write
$$
\widehat{D}_{({\frac \delta {\delta u^\gamma }})}{\frac{\widehat{\delta }}{%
\delta \widehat{u}^\beta }}=\widehat{{\Gamma }}_{\beta \gamma }^\alpha {%
\frac{\widehat{\delta }}{\delta \widehat{u}^\alpha }},
$$
where the d--isoconnection $\widehat{D}$ has the following coefficients:
$$
\widehat{{\Gamma }}{{^i}_{jk}}=\widehat{{L}}{{^i}_{jk}},\widehat{{\Gamma }}{{%
^i}_{ja}}=\widehat{{C}}{{^i}_{ja}},\widehat{{\Gamma }}{^i}_{aj}=0,\widehat{{%
\Gamma }}{^i}_{ab}=0,\eqno(7.1)
$$
$$
\widehat{{\Gamma }}{^a}_{jk}=0,\widehat{{\Gamma }}{^a}_{jb}=0,\widehat{{%
\Gamma }}{^a}_{bk}=\widehat{{L}}{^a}_{bk},\widehat{{\Gamma }}{{^a}_{bc}}=%
\widehat{{C}}{{^a}_{bc}}.
$$
The nonholonomy isocoefficients $\widehat{{w}}{{^\gamma }_{\alpha \beta }}$
are as follows:
$$
\widehat{{w}}{{^k}_{ij}}=0,\widehat{{w}}{{^k}_{aj}}=0,\widehat{{w}}{{^k}_{ia}%
}=0,\widehat{{w}}{{^k}_{ab}}=0,\widehat{{w}}{{^a}_{ij}}=\widehat{{R}}{^a}%
_{ij},
$$
$$
\widehat{{w}}{{^b}_{ai}}=-{\frac{\widehat{\partial }\widehat{{N}}{_a^b}}{%
\partial \widehat{y}^a}},\widehat{{w}}{{^b}_{ia}}={\frac{\widehat{\partial }%
\widehat{{N}}{_a^b}}{\partial \widehat{y}^a}},\widehat{{w}}{{^c}_{ab}}=0.
$$
By straightforward calculations we can obtain respectively these components
of isotorsion, ${\cal T}(\widehat{{\delta }}_\gamma ,\widehat{{\delta }}%
_\beta )={{\cal T}_{\cdot \beta \gamma }^\alpha }\widehat{{\delta }}_\alpha
, $ and isocurvature, ${\cal R}(\widehat{{\delta }}_\beta ,\widehat{{\delta }%
}_\gamma )\widehat{{\delta }}_\tau ={{\cal R}_{\beta \cdot \gamma \tau
}^{\cdot \alpha }}\widehat{{\delta }}{_\alpha },$ d--isotensors:
$$
{\cal T}_{\cdot jk}^i=\widehat{{T}}{^i}_{jk},{\cal T}_{\cdot ja}^I=\widehat{{%
C}}{^i}_{ja},{\cal T}_{\cdot ja}^I=-\widehat{{C}}{^i}_{ja},{\cal T}_{\cdot
ab}^i=0,\eqno(7.2)
$$
$$
{\cal T}_{\cdot ij}^a=\widehat{{R}}{^a}_{ij},{\cal T}_{\cdot ib}^a=-\widehat{%
{P}}{^a}_{bi},{\cal T}_{\cdot bi}^a=\widehat{{P}}{^a}_{bi},{\cal T}_{\cdot
bc}^a=\widehat{{S}}{^a}_{bc}
$$
and
$$
{\cal R}_{i\cdot kl}^{\cdot j}=\widehat{{R}}{{_j}^i}_{kl},{\cal R}_{b\cdot
kl}^{\cdot j}=0,{\cal R}_{j\cdot kl}^{\cdot a}=0,{\cal R}_{b\cdot kl}^{\cdot
a}=\widehat{{R}}_{b\cdot kl}^{\cdot a},\eqno(7.3)
$$
$$
{\cal R}_{j\cdot kd}^{\cdot i}=\widehat{{P}}{{_j}^i}_{kd},{\cal R}_{b\cdot
kd}^{\cdot a}=0,{\cal R}_{j\cdot kd}^{\cdot a}=0,{\cal R}_{b\cdot kd}^{\cdot
a}=\widehat{{P}}{{_b}^a}_{kd},
$$
$$
{\cal R}_{j\cdot dk}^{\cdot i}=-\widehat{{P}}{{_j}^i}_{kd},{\cal R}_{b\cdot
dk}^{\cdot i}=0,{\cal R}_{j\cdot dk}^{\cdot a}=0,{\cal R}_{b\cdot dk}^{\cdot
h}=-\widehat{{P}}{{_b}^a}_{kd},
$$
$$
{\cal R}_{j\cdot cd}^{\cdot i}=\widehat{{S}}{{_j}^i}_{cd},{\cal R}_{b\cdot
cd}^{\cdot i}=0,{\cal R}_{j\cdot cd}^{\cdot a}=0, {\cal R}_{b\cdot
cd}^{\cdot a}=\widehat{{S}}{{_b}^a}_{cd}
$$
(for explicit dependencies of components of isotorsions and isocurvatures on
components of d--isocon\-nec\-ti\-on see formulas (5.4) and (5.7)).

The locally adapted components ${\cal R}_{\alpha \beta }={\cal R}ic(D)(%
\widehat{{\delta }}_\alpha ,\widehat{{\delta }}_\beta )$ (we point that in
general on t--isobundles ${\cal R}_{\alpha \beta }\ne {\cal R}_{\beta \alpha
})$ of the {\bf isoricci tensor} are as follows:
$$
{\cal R}_{ij}=\widehat{{R}}{{_i}^k}_{jk},{\cal R}_{ia}=-{}^{(2)}\widehat{{P}}%
{_{ia}}=-\widehat{{P}}_{i\cdot ka}^{\cdot k}\eqno(7.4)
$$
$$
{\cal R}_{ai}={}\widehat{{P}}{_{ai}}=\widehat{{P}}_{a\cdot ib}^{\cdot b},%
{\cal R}_{ab}=\widehat{{S}}{{_a}^c}_{bc}=\widehat{S}_{ab}.
$$
For scalar curvature, $\overleftarrow{{\ R}}=Sc(\widehat{D})=\widehat{G}%
^{\alpha \beta }\widehat{R}_{\alpha \beta },$ we have
$$
Sc(\widehat{D})=\widehat{R}+\widehat{S},\eqno(7.5)
$$
where $\widehat{R}=\widehat{g}^{ij}{\widehat{R}}_{ij}$ and $\widehat{S}=%
\widehat{h}^{ab}\widehat{S}_{ab}.$

The {\bf isoeinstein--isocartan equations with prescribed
N--iso\-con\-nec\-ti\-on} and
h(hh)-- and v(vv)--isotorsions on v--isobundles (compare with isoeinstein
iso\-equa\-ti\-ons (2.11)) are written as
$$
\widehat{R}^{\alpha \beta }-\frac 12\widehat{g}^{\alpha \beta }(%
\overleftarrow{R}+\widehat{\Theta }-\lambda )={\kappa }_1(\widehat{t}%
^{\alpha \beta }-\widehat{\tau }^{\alpha \beta }),\eqno(7.6)
$$
and
$$
\widehat{T}_{\cdot \beta \gamma }^\alpha +{{G_\beta }^\alpha }\widehat{{T}}{{%
^\tau }_{\gamma \tau }}-{{G_\gamma }^\alpha }\widehat{{T}}{{^\tau }_{\beta
\tau }}={\kappa }_2\widehat{{Q}}{^\alpha }_{\beta \gamma },\eqno(7.7)
$$
where $\widehat{{Q}}{_{\beta \gamma }^\alpha }$ spin--density of matter
d--isotensors on locally anisotropic and homogeneous
 isospace, ${\kappa }_1$ and ${\kappa }_2$ are the
corresponding interaction constants and ${\lambda }$ is the cosmological
constant, $\widehat{t}^{\alpha \beta }$ is a source isotensor and $\widehat{%
\tau }^{\alpha \beta }$ is the stress--energy isotensor and there is
satisfied the generalized Freud isoidentity
$$
\widehat{G}_{~\beta }^\alpha -\frac 12\delta _{~\beta }^\alpha (\widehat{%
\Theta }-\lambda )=\widehat{U}_{~\beta }^\alpha +\widehat{\delta }_\rho
\widehat{V}_{\quad \beta }^{\alpha \rho },\eqno(7.8)
$$
where
$$
\widehat{G}_{~\beta }^\alpha =\widehat{R}_{\quad \beta }^\alpha -\frac
12\delta _{~\beta }^\alpha \overleftarrow{R},
$$
$$
\widehat{U}_{~\beta }^\alpha =-\frac 12\frac{\widehat{\delta }\widehat{%
\Theta }}{\widehat{\partial }(\widehat{D}_\alpha \widehat{g}^{\gamma \delta
})}\widehat{D}_\beta \widehat{g}^{\gamma \delta }
$$
and
$$
\widehat{V}_{\quad \beta }^{\alpha \rho }=\frac 12[\widehat{g}^{\gamma
\delta }\left( \delta _{~\beta }^\alpha \widehat{\Gamma }_{\alpha \delta
}^\rho -\delta _{~\delta }^\alpha \widehat{\Gamma }_{\alpha \beta }^\rho
\right) +%
$$
$$
\widehat{g}^{\rho \gamma }\widehat{\Gamma }_{\beta \gamma }^\alpha -\widehat{%
g}^{\alpha \gamma }\widehat{\Gamma }_{\beta \gamma }^\rho +\left( \delta
_{~\beta }^\rho \widehat{g}^{\alpha \gamma }-\delta _{~\beta }^\alpha
\widehat{g}^{\rho \gamma }\right) \widehat{\Gamma }_{\gamma \rho }^\rho ].
$$

By using decompositions (7.1)--(7.5) it is possible an explicit projection
of equations (7.6)--(7.8) into vertical and horizontal isocomponents (for
simplicity we omit such formulas in this work).

Equations (7.6) constitute the fundamental field equations of Santilli
 isogravitation  \cite{Santilli1978,Santilli1979,Santilli1980,Santilli1996}
 written in this case for vector isobundles provided with
 compatible N- and d--isoconnection and isometric structures.
The algebraic equations (7.7) have been here added, apparently
 for the first    time for isogravity with isotorsion (see also
 \cite{VacaruGoncharenko,VacaruCL,VacaruSupergravity}
for locally anisotropic gravity and supergravity)
 in order to close the system of
gravitational isofield equations (really we have also to take into account
the system of constraints (4.5) if locally anisotropic inhomogeneous
 gravitational
isofield is associated to a d--isometric (4.6), or to a d--isometric (6.5)
if the isogravity is modelled on a tangent isobundle). It should be noted
here that the system of isogravitational field equations (7.8)
 presents a synthesis for
vector isobundles of equations introduced by Anastasiei, \cite{Anastasiei}
and \cite{MironAnastasiei}, and of equations (2.11) and (2.12) considered in
the Santilli isotheory.

We note that on la--isospaces the divergence
$$
D_\alpha [\widehat{G}_{~\beta }^\alpha -\frac 12\delta _{~\beta }^\alpha (%
\widehat{\Theta }-\lambda )]=\widehat{U}_\beta \eqno(7.9)
$$
does not vanish (this is a consequence of generalized isobianchi (5.8), or
(5.9), and isoricci isoidentities (5.11), or (5.12)). The problem of
nonvanishing of such divergences for gravitational models on vector bundles
provided with nonlinear connection structures was analyzed in \cite
{Anastasiei} and \cite{MironAnastasiei}.

The problem of total conservation laws on isospaces has been studied
in detail in ref. \cite{Santilli1978} by reformulating  all isospaces
considered in that paper in terms  of the isominkowskian space,
 with consequential elimination of curvature  which permits the
construction of a  universal symmetry and related total conservation
 laws for all possible isometric.

The latter studies concerning vector isobundles with
 N--isoconnections will be considered in some future works.

We end this subsection by emphasizing that isofield equations of ty\-pe
(7.6)--(7.8) can be similarly introduced for the particular cases of
lo\-cal\-ly anisotropic isospaces with metric (6.5) on $\tilde {TM}$ with
coefficients pa\-ra\-met\-ri\-zed as for the isolagrange, (6.3), or
 isofinsler, (6.4), isospaces.

\section{Concluding Remarks and Further Possibilities}

One of the most important aspects we attempted to convey in this work is
the possiblity to formulate isotopic variants of extended Finsler geometry
and the application of this isogeometric background in contemporary theoretical
 and mathematical physics. The approach adopted here provides us an
essentially self--contained, concise and significantly simple treatment
of the material on bundle isospaces enabled with compatible isotopic nonlinear
 and distinguished isoconnections and isometric structures.

A remarkable features worth recalling is that the considerable broadening of
 the capabilities of the isotheory via the additional of nonlinear,
 nonlocal and noncanonical effects, is done via the same abstract
 axioms of the conventional formulations.

In this paper we have discussed the basic geometric constructions for
 isotopic spaces with inhomogeneity and local anisotropy.
 We have computed the distinguished
 isotorsions and isocurvatures. It was shown how to write a manifestly
isotopic model of gravity with locally anisotropic
 and inhomogeneous interactions of isofields.
The assumptions made in deriving the results are similar to those for
 the geometry of isomanifolds and to the isofield theory.

There are various possible developments of the ideas presented here. One of
the necessary steps is the definition of locally anisotropic and
 inhomogeneous isotopic spinors
 and explicit constructions of physical models with  isospinor, isogauge and
 isogravitational interactions on locally anisotropic isospaces. The
 problem of formulation of conservation laws on
 locally anisotropic and inhomogeneous isospaces and for
  locally anisotropic and inhomogeneous
isofield interactions presents a substantial intrested for investigations.
  Here we add
 the theory of isostochastic processes, the supersymmetric extension of
 the concept of isotheory as well possible generalizations of the mentioned
 constructions for higher order anisotropies in string theories. These
tasks remain for future research.

\section*{Acknowledgement}

The author wish to express generic thanks to the referees for a detailed
control and numerous constructive suggestions.

\end{document}